\RequirePackage{fix-cm}
\documentclass[twocolumn,final]{svjour3}              
\usepackage{graphicx}
\usepackage {epsfig}
\usepackage{epstopdf}
\usepackage{xcolor}
\usepackage{tikz}
\usepackage{subcaption}
\usetikzlibrary{decorations.text,shapes}
\journalname{Journal of Materials Science}
\usepackage[hidelinks]{hyperref}                        
\usepackage{cite}
\usepackage[T1]{fontenc}
\usepackage{filecontents}
\newcommand{\subparagraph}{}
\usepackage[compact]{titlesec}
\titleformat{\paragraph}[runin]{\normalfont\normalsize\bfseries}{\theparagraph}{1em}{}

\begin{document}
\title{Review of the Synergies Between Computational Modeling and Experimental Characterization of Materials Across Length Scales}

\titlerunning{Modeling and Experimental Characterization Across Length Scales}        

\author{R\'emi Dingreville  \and
        Richard A. Karnesky \and
        Guillaume Puel \and
        Jean-Hubert Schmitt
}

\institute{R. Dingreville \at
              Sandia National Laboratories, Albuquerque, NM 87185, USA\\
            \email{rdingre@sandia.gov}
           \and
           R. Karnesky \at
              Sandia National Laboratories, Livermore, CA 94550, USA\\
            \email{rakarne@sandia.gov}
           \and
           G. Puel, J.-H. Schmitt \at
            Laboratoire M\'ecanique des Sols, Structures et Mat\'eriaux (MSSMat), CNRS UMR 8579, Centrale/Sup\'elec, Universit\'e Paris-Saclay, 92290 Ch\^atenay Malabry, France 
}

\date{Received: 11 September 2015 / Accepted: date}

\maketitle

\begin{abstract}
With the increasing interplay between experimental and computational approaches at multiple length scales, new research directions are emerging in materials science and computational mechanics. Such cooperative interactions find many applications in the development, characterization and design of complex material systems. This manuscript provides a broad and comprehensive overview of recent trends where predictive modeling capabilities are developed in conjunction with experiments and advanced characterization to gain a greater insight into structure-properties relationships and study various physical phenomena and mechanisms. The focus of this review is on the intersections of multiscale materials experiments and modeling relevant to the materials mechanics community. After a general discussion on the perspective from various communities, the article focuses on the latest experimental and theoretical opportunities. Emphasis is given to the role of experiments in multiscale models, including insights into how computations can be used as discovery tools for materials engineering, rather than to ``simply'' support experimental work. This is illustrated by examples from several application areas on structural materials. This manuscript ends with a discussion on some problems and open scientific questions that are being explored in order to advance this relatively new field of research.  
\end{abstract}
\section{Introduction}
\label{sec:intro}
Recent advances in theoretical and numerical methods, coupled with an increase of available high performance computing resources, have led to the development of large-scale simulations in both Materials Science and Mechanical Engineering, with the goal of better characterization and optimization of materials performance. These advanced computational capabilities extend over a large span of length scales---ranging from the atomistic to the continuum scale---and serve as investigative tools to get a greater insight into structure-properties relationships.
As these predictive modeling capabilities become more comprehensive and quantitative, a comparable level of details from experiments and characterization tools is now not only necessary to validate these multiscale models, but also to motivate further theoretical and computational advances. The present complexity of the multiscale approach leads also to the need for new algorithms to bridge length scales and for model reduction. Strong coupling between experimental data and numerical simulation is required for predicting macroscopic behavior with a fine description of local fields (i\@.e\@.~mechanical, compositional, electrical, etc.).
An illustration of such synergies is the coupling of experimental tools such as Transmission Electron Microscopy (TEM) and synchrotron-based X-ray microscopy that enable the collection of microstructural statistical data comparable to the output of three-dimensional (3-D) Crystal Plasticity Finite Element Method (CPFEM)-based simulations~\cite{Ohashi09IJP,Zaafarani06ActaMat}.

\begin{figure*}[!htb]
  \centering
    \begin{subfigure}[t]{0.49\linewidth}
      \captionsetup{width=0.9\textwidth}
      \centering
        \includegraphics[width=\linewidth]{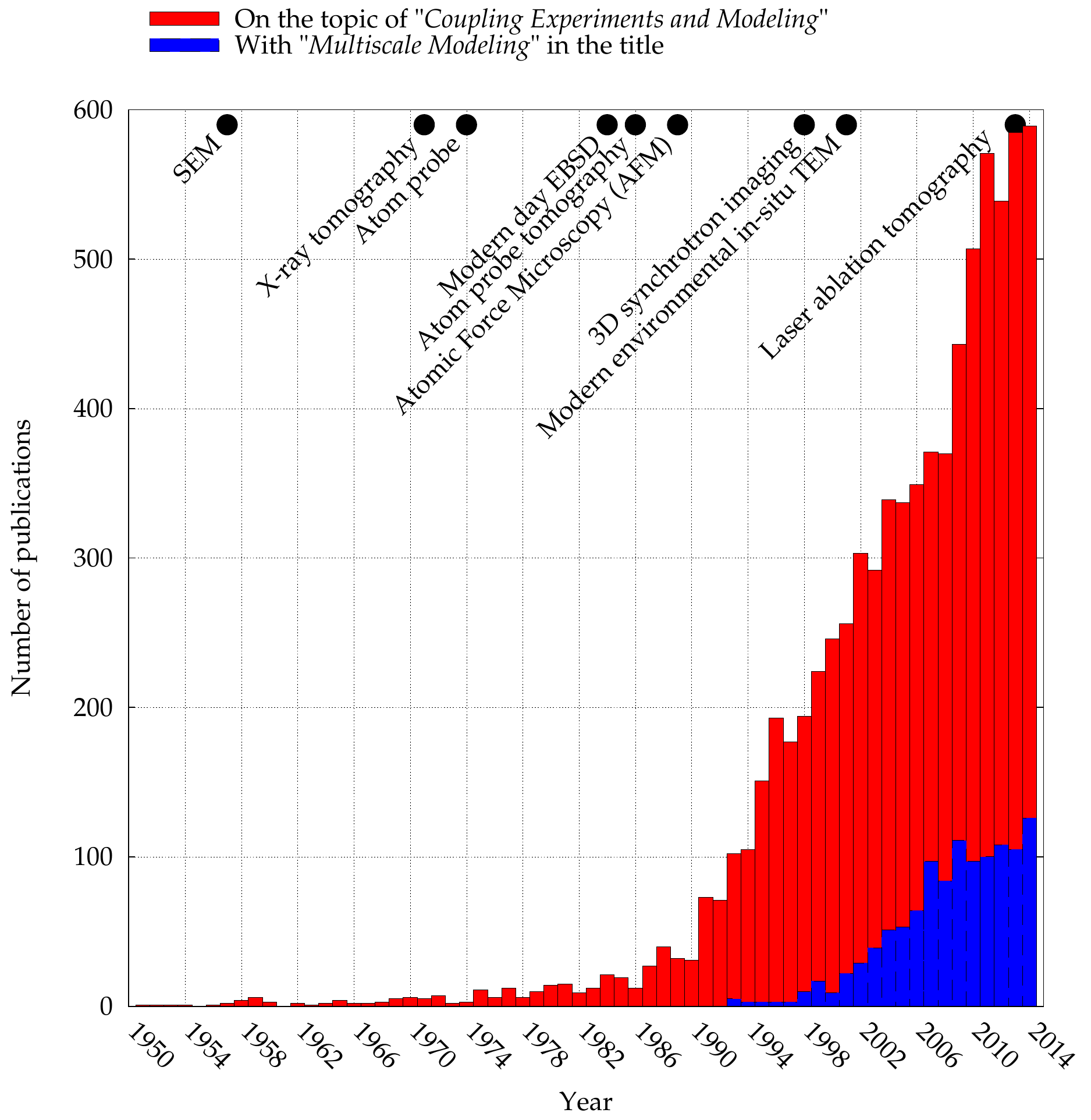}
        \caption{Papers with ``\emph{Coupling Experiments and Modeling}'' as topic area and with ``\emph{Multiscale Modeling}'' in the title. The circle symbols denote the introduction of specific experimental technologies.\label{fig:WOSa}}
    \end{subfigure}
    \begin{subfigure}[t]{0.49\linewidth}
      \captionsetup{width=0.9\textwidth}
      \centering
        \includegraphics[width=\linewidth]{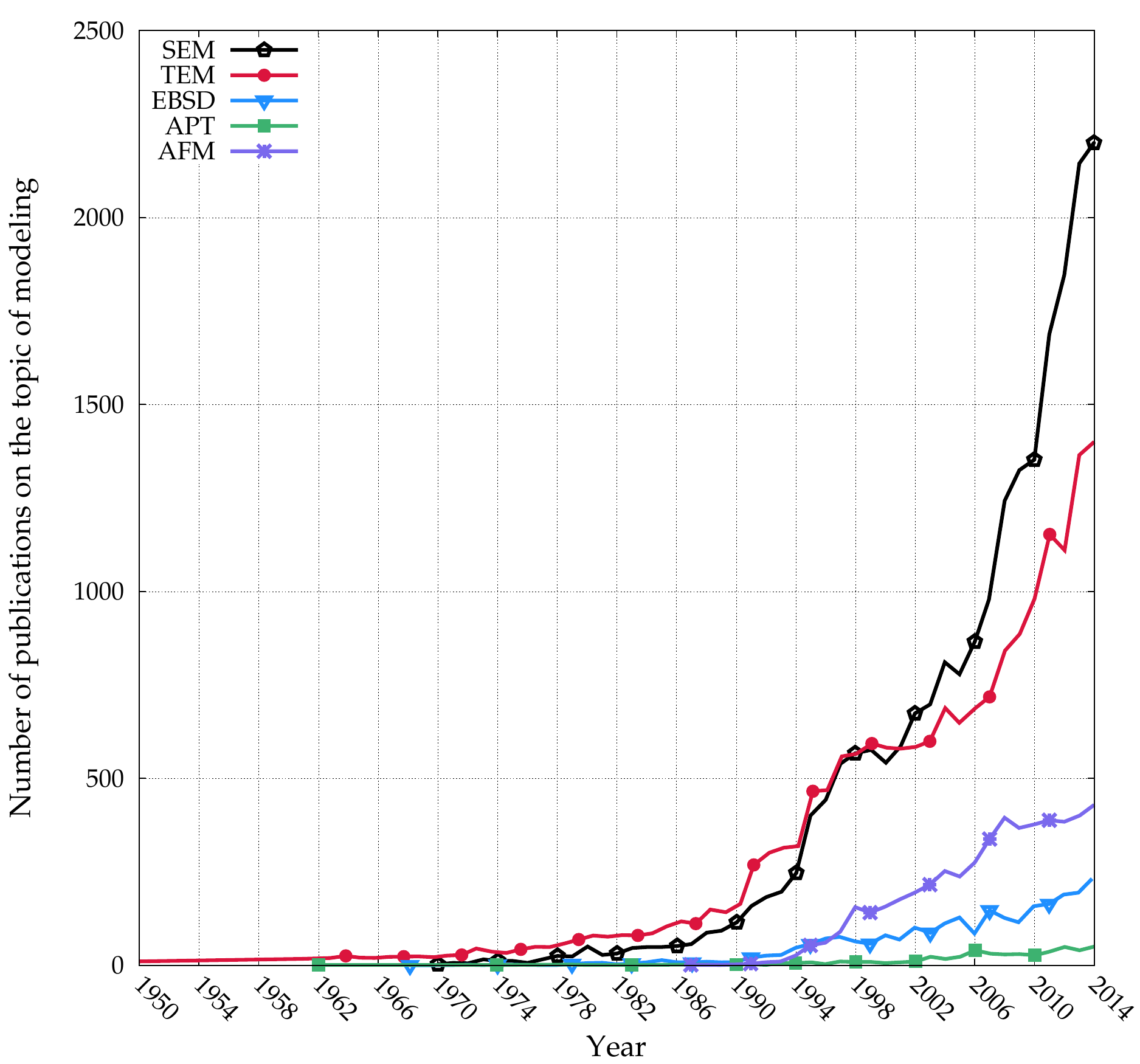}
        \caption{Papers focused on modeling and various experimental techniques.\label{fig:WOSb}}
    \end{subfigure}
    \caption{Evolution of the number of journal articles published in the fields of \emph{Materials Science} and \emph{Mechanics} according to ISI Web of Knowledge.}\label{fig:WOS}
\end{figure*}

As illustrated in Fig.~\ref{fig:WOS}, based on a citation report extracted from \href{http://webofknowledge.com}{ISI Web of Knowledge}, there is an increasing trend of articles in the fields of both \emph{Materials Science} and \emph{Mechanics} to report on both multiscale modeling and experiments. In Fig.~\ref{fig:WOSa}, we consider those that denote a topic area of ``\emph{Coupling Experiments and Modeling}''. A similar trend is observed for articles containing ``\emph{Multiscale Modeling}'' in the title for research areas limited to the same fields of \emph{Materials Science} and \emph{Mechanics}. Both progressions in the number of journal articles published highlight the increasing cross-pollinations between the modeling community and the experimental community. While the term ``\emph{coupling}'' is far too inclusive to extract any real correlation between the emergence of advanced characterization methodologies and growth of research into ``\emph{Coupling Experiments and Modeling}'', it is interesting to put in perspective this increase of publications with respect to the successive introduction of novel experimental characterization methodologies such as Scanning Electron Microscopy (SEM)~\cite{VonArdenne1938ZFP}, Atom Probe Tomography (APT)~\cite{Seidman2007}, SEM-based Electron Back Scattered Diffraction (EBSD)~\cite{Venables73PhilMag} and 3-D high energy synchrotron X-ray approaches~\cite{Poulsen2004}. As such, the increasing use of various individual experimental techniques within modeling paradigms is illustrated in Fig.~\ref{fig:WOSb}.
%
%
\subsection{Background}
Just what is this emerging field? After all, many multiscale modeling strategies discussed in the mechanics and Materials Science communities couple modeling and experiments intrinsically, i\@.e\@., they involve models of different physics at multiple scales and experimentally validate those models at the corresponding length/time scales.
But because of the interdisciplinary nature of the fields, a novel branch in Materials Science that relies on the synergy between experimental and computational approaches at multiple length-scales is opening up new frontiers and research directions at the crossroads of both traditional Computational and Experimental Materials Science, in addition to the emerging field of Integrated Computational Materials Engineering (ICME).
Such cooperative interactions find many applications in the development, the characterization and the design of complex material systems. Yet more possibilities are waiting to be explored and many new questions of physical, numerical, analytical or characterization nature have materialized. These things all make this emerging field an extremely fruitful and promising area of research.

Almost all problems in Materials Science are multiscale in nature: fundamental small scale physical mechanisms and processes occurring at the atomic length scale and femtosecond time scale have a profound impact on how materials perform at larger spatial and time scales.
Despite this inherent multiscale character, effective macroscopic models and measurements are used---often with satisfactory accuracy---that account intrinsically for the effects of the underlying microscopic processes. For example, a conventional tensile test directly measures the ultimate tensile strength, the maximum elongation, the reduction in area, and determines a phenomenological uniaxial stress-strain curve. Such an experiment can be used effectively to calibrate a Hollomon-type power law relation between the stress and the amount of plastic strain regardless of the fundamental microstructural mechanisms responsible for the macroscopic response.

However, the above class of macroscopic models and experiments have some limitations. The first obvious limitation is the complete neglect of microscopic mechanisms that are sometimes of interest. In our tensile test example, we may want to know how changes in microstructure would lead to a change in the mechanical response. It is often of interest to know and characterize the microstructural defects such as dislocations, martensite formation, and mechanical twinning that are at the origin of plastic deformation, not just the macroscopic flow. Such information requires modeling and experimental tools that can track such defects. These include, for example, CPFEM or discrete dislocation dynamics (DDD) on the modeling side and TEM, X-ray tomography or SEM-based EBSD on the experimental side.
Another limitation is associated with the heuristic nature of the macroscopic models and experimental measurements. These effective models often have an empirical foundation limiting their predictive capabilities, while the interpretation of experimental data may be challenging in the case of complex material systems subjected to the intricate interplays of thermodynamics, transport (diffusion and phase transformations) and non-linear behavior. As we will describe later, one of the advantages of detailed multiscale models is their ability to switch which physical mechanisms are and are not active in the material to decouple and examine the importance of different mechanisms and help in the interpretation of experimental data.
Finally, another limitation is that it is difficult to gain insight into the margins and uncertainty due not only to test-to-test variation, but also the natural stochasticity at the lower length scales of engineering materials that makes an ``exact'' knowledge of the underlying microstructure and active mechanisms impossible.

For these reasons, one might be tempted to switch completely to a detailed microscopic characterization (either model-wise, experimentally or both) that has the highest resolution and strongest physical foundation. However, this is not always a favorable strategy; not only because the microscopic models and experimental setup are often difficult to handle, but also because it requires complex procedures to extract the information of interest and perform meaningful data mining.

This is where the coupling between experiments and simulations is essential. By combining multiscale modeling and corresponding experiments, one hopes to take advantage of the simplicity, efficiency and mechanistic insight gained from the models, as well as the physicality, meaningful reproducibility and ``reality check'' provided by the experiments.
%
%
	\subsection{Different communities, different perspectives}
Various perspectives from diverse communities lead to different combinations of experiments and simulations.

\paragraph{Perspective of Solid Mechanics:} As pointed out by Hortstemeyer~\cite{Horstemeyer2012}, the Solids Mechanics community tends to adapt a ``\emph{top-down}'' or ``\emph{downscaling}'' approach. Such tendency is driven by the need to characterize materials performance through the use of internal state variables (ISVs) and degrees of freedom (DoF) representing an underlying microstructure at the pertinent length scales. Stemming for the most part from the Micromechanics community~\cite{Eshelby57, mura1987micromechanics, suquet1987elements}, this led to the development of hierarchical and nested methods in which both numerical/theoretical models and experiments probe properties and mechanisms down to various subscales to characterize the evolution of ISVs and thus describe the materials behavior at the macroscale. For example, a large body of work is currently being conducted on the quantification and characterization of the behavior of small size systems (i\@.e\@. when the structure size compares with the microstructure size) such as micro- and nano- electromechanical systems (MEMS and NEMS)~\cite{ke2005experiments, pelesko2002modeling, cao2009molecular, dorogin2013real} or 3-D printing structures with thin walls and beams~\cite{zaumseil2003three, pan2012hierarchical, vaezi2013review}.

\paragraph{Perspective of Materials Science:} The Materials Science community focuses generally on a ``\emph{bottom-up}'' or ``\emph{upscaling}'' approach. Indeed, the basis of Materials Science involves studying the structure of materials and their associated length scales, and relating them to their properties. While Materials Science researchers (notably those in the Mechanics community, who study the interplay between microstructure and mechanical response) have always connected theory and experiments, it was not until the mid-nineties and the emergence of advanced characterization tools (see Fig.~\ref{fig:WOS}) that multiscale studies have proliferated to investigate structure-property relations. Synergies between experiments and modeling emerging from this community are motivated by the rationale that the different structures composing a material (e\@.g\@. grain size and texture, dislocation network, inclusions) dictate its performance properties. For example, many studies now focus on the characterization of the microstructural evolutions through \emph{in situ} observations~\cite{legros2008situ, wang2011situ} or on 3-D statistical descriptions of microstructure~\cite{rollett2007three, groeber2008framework, groeber2014dream}.

\paragraph{Perspective of Applied Mathematics:} The above-mentioned communities require the proven methodologies of the Applied Mathematics community~\cite{brandt2002multiscale, E07} supporting the characterization and study of materials behavior across multiple length scales both from the modeling and experimental aspects. The major focal points of this community lies in the scale bridging. Thus, methodologies such as dimensional analysis~\cite{barenblatt1996scaling, cheng2004scaling}, fractal analysis~\cite{miguel2001intermittent, Csikor12102007, chen2010bending}, statistical analysis~\cite{bai2005statistical, liu2013computational}, graph theory~\cite{bachmann2011grain, germain2014identification}, parallel computing~\cite{bjorstad1986iterative, pothen1990partitioning, karypis1998fast}, and model reduction~\cite{krysl2001dimensional, willcox2002balanced} started from the applied mathematics community and were successfully used by both modelers and experimentalists in the Materials Science and Solid Mechanics communities.

\paragraph{} Although the combination of experiments and simulations is common to many disciplines, this review will focus on recent progress made in metals. The authors readily recognize many areas where the integration of experiments with multiscale modeling has affected the Materials Science community, for example in the area of polymers~\cite{bouvard2009review, li2013challenges}, or biomaterials~\cite{de2007multiscale, praprotnik2008multiscale, murtola2009multiscale}.
The purpose of this manuscript is to provide the Materials Science community with a coherent overview of the status of recent progress made into how Computational Materials Science can be exploited as discovery tools for materials engineering and directly integrated with experimental work. It is our hope that this summary will help the reader to understand (i) the recent successes of predicting various physical phenomena and mechanisms in materials systems enabled by the collaboration between experimentalists and modelers; (ii) the main obstacles that one has to face when experiments are integrated into various computational mechanics frameworks; and (iii) a perspective on the upcoming issues to solve in order to make this emerging field of Materials Science a more powerful and predictive approach.
The manuscript is organized as follows. Section~\ref{sec:recent} highlights recent synergies at various scales ranging from the atomistic scale to the continuum scale from an experimental perspective. Section~\ref{sec:models} provides a discussion on the recent developments of models and upscaling strategies between the various length scales. Section~\ref{sec:roles} discusses the roles of experiments in multiscale models.  The review ends with a discussion on some problems that have to be addressed in order to successfully leverage the experiment-modeling synergy.
\section{Recent evolutions in experimental techniques across multiple length scales}
\label{sec:recent}
The quantitative description of microstructure is an essential requirement for multiscale models describing the macroscopic properties of materials~\cite{DeHoff1968}. Thus, the rapid evolution of experimental characterization tools opens up new opportunities to measure a wide variety of local fields accurately within a large volume and subsequently incorporate these measurements into a model at the corresponding scale. Not only can this type of experimental data be used to calibrate models, it can also drive the development and improvement of new theoretical approaches.

Recent developments of experimental instrumentation including (i) new capabilities for mechanical measurements and microstructural analyses at the nanometer length scale (Section~\ref{sec:atomiclevel}); (ii) the development of full field measurements for different parameters beyond displacement field quantification (Section~\ref{sec:fieldmeasurements}); and (iii) the improvement of 3-D characterizations (e\@.g\@. tomography; Section~\ref{sec:3Dmeasurements}), have a major impact on the modeling community.
%
%
\subsection{Nanoscale microstructural and mechanical analysis}
\label{sec:atomiclevel}
The advent of optimized materials systems engineered at the near-atomic scale requires the quantification of microstructural features (local composition, crystalline structure) and local mechanical measurements at that same scale.
%
%
\paragraph{Local chemical composition measurements:} Over the years, a large number of small-scale measurement techniques have been developed to quantify the local chemical composition in structures through recent advances in analytical High-Resolution (HR)TEM~\cite{Fultz2012, den2013estimation} and in analytical Scanning (S)TEM.
For instance, new STEM systems employ very high sensitivity X-ray detectors~\cite{Watanabe1999, Genc2009} allowing for very sensitive 2-D and 3-D measurement of composition.
Coupled with
electron energy-loss spectroscopy
and
high-angle annular dark-field imaging, it is now possible to reach a spatial accuracy of few atoms, allowing for the local identification of the chemical composition and, in specific cases, the oxidation state~\cite{Zhou2012}.
However, the acquisition of 3-D characterization of the composition within a TEM remains challenging. For this, the Materials Science community turns to APT~\cite{Zhou2012}.

\begin{figure}[!htb]
\begin{center}
	\includegraphics[width=7.25cm]{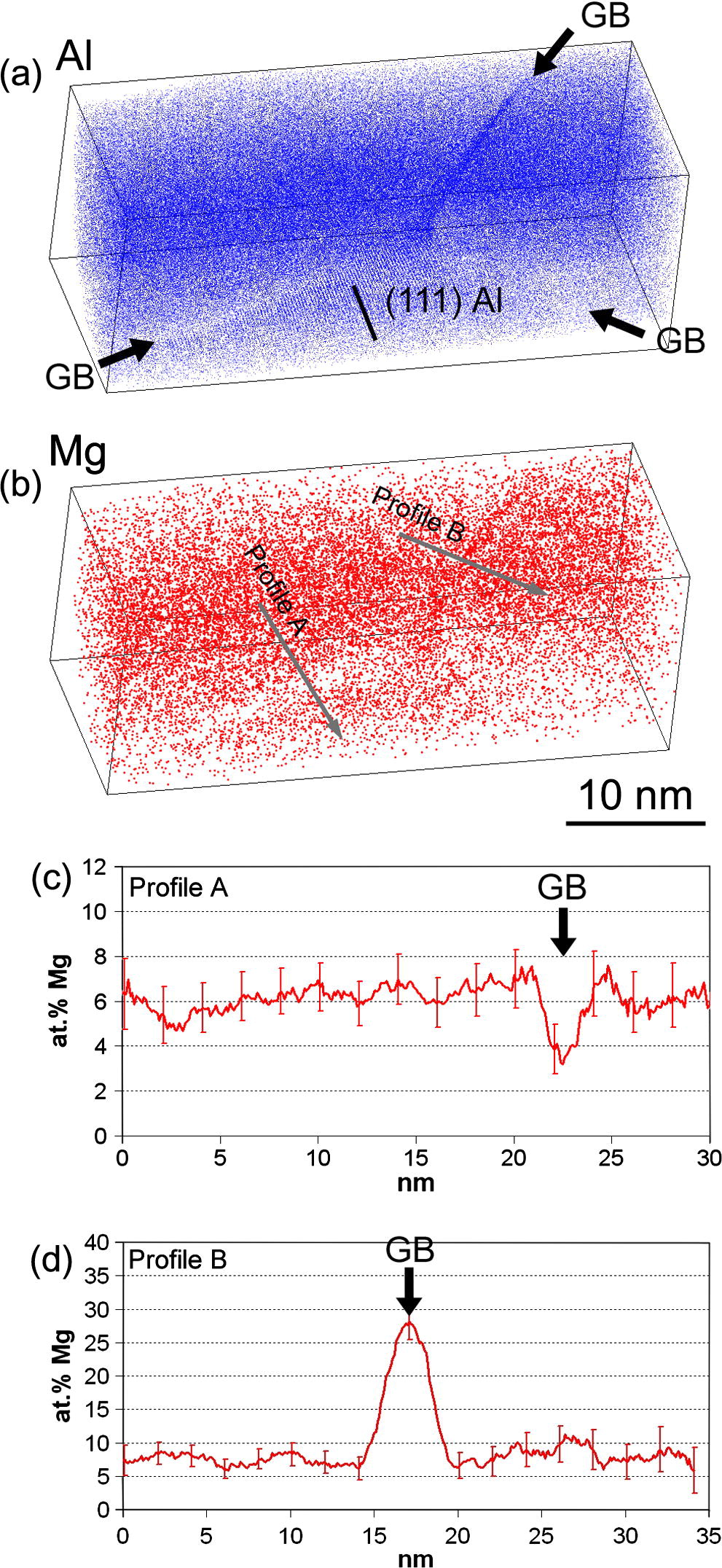}
	\caption{Atom probe analysis of Mg concentration at a grain boundary (GB) triple junction in a Al--Mg alloy deformed at 200$^\circ$C\@: (a) Al atoms; (b) Mg map with indication of the two concentration profiles shown in (c) for Profile A, and (d) for Profile B\@. Depletion and segregation of Mg is quantified close to the GBs within a 2.5~nm step size~(the sampling volume thickness is 1~nm). Reprinted from Ref.~\cite{Sauvage2014}, Copyright~(2014), with permission from Elsevier.}\label{fig:AtomProbe}
\end{center}
\end{figure}

Even though it is exercised by a small community, APT~\cite{Blavette1993, Seidman2007, Marquis2013, Marquis2014} has shown the most progress in improved resolution of small length scale compositional measurements. The technique enables the identification of the chemical nature of atoms using time-of-flight mass spectrometry and, at the same time, their positions within the sample with sub-nanometer resolution. Data obtained from APT provide new insights about mechanisms controlling the evolution of local chemical composition.
Figure~\ref{fig:AtomProbe} exemplifies a recent study of grain boundary segregation in an Al--Mg alloy~\cite{Sauvage2014} using APT to reveal the mechanisms controlling dynamic precipitation and segregation during severe plastic deformation of aluminum alloys.
Due to the relatively small dimensions ($\sim10^6~\textrm{nm}^3$) of the materials volumes and the reasonable number of atoms ($\sim10^7$~atoms) that can be characterized using this technique, an increasing number of studies couple atomistic simulations and APT (ISI Web of Knowledge citation report extracted 460 articles between 2002 and 2015, see Fig.~\ref{fig:WOSb}).
%
%
%
\paragraph{\emph{In situ} mechanical field measurements:} Experimental techniques being used to measure the elastic fields at the nanometer length scale are now also available. Among them, high resolution X-Ray Diffraction (XRD)~\cite{Bernier2011, reischig2013advances, Maire2014, Schuren14} is particularly interesting since it allows for lattice orientation, and lattice strain tensor coordinates to be measured with great accuracy. Because the full lattice strain tensors are available, corresponding mean stress tensors may be calculated directly from the experimental data at the sub-grain level unambiguously using classical elasticity theories.
Such measurements are especially useful to study strain/stress evolution \emph{in situ} and conduct residual stress analysis. They are well-suited to validate polycrystalline models such as CPFEM~\cite{van2013deformation, obstalecki2014quantitative, lentz2015situ} or to inform local mechanistic criteria such as crack emission or damage at grain boundary, for example. Thus, Van Petegem \emph{et al.}~\cite{van2013deformation} conducted \emph{in situ} mechanical testing using XRD on nanocrystalline Ni to observe the evolution of inter-granular stresses and used molecular dynamics (MD) simulations and crystal plasticity models to help with the interpretation of the experimental results. Similarly, recent femtosecond XRD experiments have been combined with atomistic simulations to characterize lattice dynamics in shock-compressed materials to study the ultrafast evolution of microstructure~\cite{clark2013ultrafast, milathianaki2013femtosecond}.
%
%
%
\paragraph{Mechanical properties measurements at small scales:} Motivated by the development of nanoscale systems (e\@.g\@. NEMS) and by the explosion of sub-grain level simulations (21,684 publications reported from ISI Web of Knowledge for the 2010--2015 time period in the research area solely limited to the field of \emph{Materials Science} on the topic of ``\emph{Molecular Dynamics}'', for example), mechanical testing at the micro-scale is currently a very popular field of research.

A conventional approach to obtain local mechanical properties is through hardness measurements. Some development in modeling the nanoindentation process is given in Section~\ref{sec:multiphysics}, but sufficed to say that this popular experimental technique (8,952 publications reported from ISI Web of Knowledge for the 2010--2015 time period in the research area solely limited to \emph{Materials Science}) enables the extraction of various quantitative characteristics of the materials being tested from the load-displacement curve recorded.
These include, for example, the elastic behavior (both during loading and unloading), macroscopic yield stress, hardness, local contact characteristics, and internal stresses in near-surface layers~\cite{Golovin2008}. The magnitude of the indentation force applied and the shape of the indenter change not only the characteristic size of a locally deformed region but also the contributions of the elastic and plastic deformations and in the sub-surface active deformation mechanisms. Therefore, when coupled with such experiments, various types of computational models are selected depending on the relative deformation mechanisms activated:
\begin{itemize}
	\item At the initial stages of the indenter penetration, when the nucleation and migration of individual structural defects come into action, atomistic simulations are used to mimic nanoindentation experiments. These support the analysis and interpretation of the experimental results to gain insight on the deformation mechanisms~\cite{lodes2011influence, begau2011atomistic, ruestes2014atomistic}. For example, Lodes \emph{et al.}~\cite{lodes2011influence} studied the indentation size effect and pop-in behavior in CaF$_2$ single crystals, using both nanoindentation experiments and MD simulations.
	\item For later stages of indenter penetration when  considerable plastic deformation occurs, atomistic tools are no longer appropriate, rather CPFEM-type calculations become more relevant simulation tools~\cite{zambaldi2012orientation, selvarajou2014orientation}. For example, by combining microindentation with a CPFEM approach, Yao ~\emph{et al.} were able to quantify the plastic parameter of slip systems in tungsten single crystal~\cite{Yao2014}. For polycrystalline materials, local behavior measured by nanoindentation was combined with a corresponding CPFEM model to study the anisotropic response of $\alpha$-titanium~\cite{zambaldi2012orientation}.
\end{itemize}
\emph{In situ} nano-hardness experiments within a TEM~\cite{Minor2001, legros2010quantitative, kiener2011situ, ohmura2012effects} consisting of the integration of nanoindentation with real-time electron imaging also merit mentioning. Classically, the main objective of these experiments is real-time study of the interplay between deformation mechanisms and the mechanical response at the nanometer length scale. Such experiments are computationally well-suited for atomistic modeling simulations~\cite{issa2015situ, liu2015situ} or discrete dislocation dynamics simulations~\cite{soer2004effects, Gouldstone2007, bufford2014situ}, where these computational paradigms are used to provide new insights on the fundamental physical phenomena dominating dominating the mechanical response. For example,  Bufford \emph{et al.}~\cite{bufford2014situ} used \emph{in situ} nanoindentation to study  plasticity and work hardening in aluminium with incoherent twin boundaries. Discrete dislocation dynamics simulations were used to clarify the role of stacking-fault energy in the difference of deformation mechanisms between nanotwinned copper and twinned aluminium.

Novel experimental advances at the nanoscale allow for a precise insight of the compositional, microstructural and mechanical states of materials. As indicated by multiple examples above, this type of characterization is of primary importance to obtain quantitative information for the analysis of active mechanisms (e\@.g\@. the plastic behavior and defect activities near a grain boundary). However, in order to obtain accurate and reliable measurements at this length scale, numerical simulations have to be consistent with this scale to identify relevant mechanisms, or obtain materials properties by reverse modeling.
%
%
\subsection{Local field measurements and analysis}\label{sec:fieldmeasurements}
Recent experimental advances in local field measurements is another area of Materials Science where experiments have been impactful within the modeling community since such experiments have enabled a higher degree of verification and validation of models beyond the macroscopic and average behavior.

%
%
\paragraph{Surface displacement fields from microgrid techniques:} Although not very accurate and with a low resolution, photoelasticimetry and holographic interferometry are simple and useful legacy tools to visualize the displacement fields at surfaces. For example, in 1987, Boehler and Raclin~\cite{Boehler1987}  leveraged the natural roughness of a tensile sample to develop a technique of pseudo relief in order to deduce from two subsequent images a complete displacement field on the surface.
Since then, refinements in the resolution and visualization of the surface displacement fields came from printing a regular microgrid on tensile SEM samples for \textit{in situ} experiments~\cite{Rey1984, Allais1994, Rey1997}.
Initially used for deep drawing tests, millimeter circles were printed on the surface of the sample before deformation to (i) measure the direction of the principal deformation axes and the principal strains and (ii) visualize the occurrence of the necking in a complex stamped part. The microgrids were usually made out of two orthogonal sets of parallel lines of sputtered metal onto the sample surface. The thickness of the lines was a few tenths of a micrometer and mesh size was a few square micrometers.
One of the advantages of using regular grids like this is easy detection of shear localizations~\cite{Martin2013}. More recently such a technique has been coupled with CPFEM simulations~\cite{Raphanel2000, soppa2001experimental, Heripre07} in order to understand the correlations between the mechanical behavior and the underlying microstructure. CPFEM calculations are carried out on a mesh of the microstructure based on orientation imaging microscopy (generally EBSD mapping) and the simulation results are compared to the experimental local strain field measurements obtained from digital image correlation of the microgrid. Few attempts have been made so far to get similar in-volume information (see Section~\ref{sec:3Dmeasurements}). 

\begin{figure}[!htb]
\begin{center}
	\includegraphics[width=8cm]{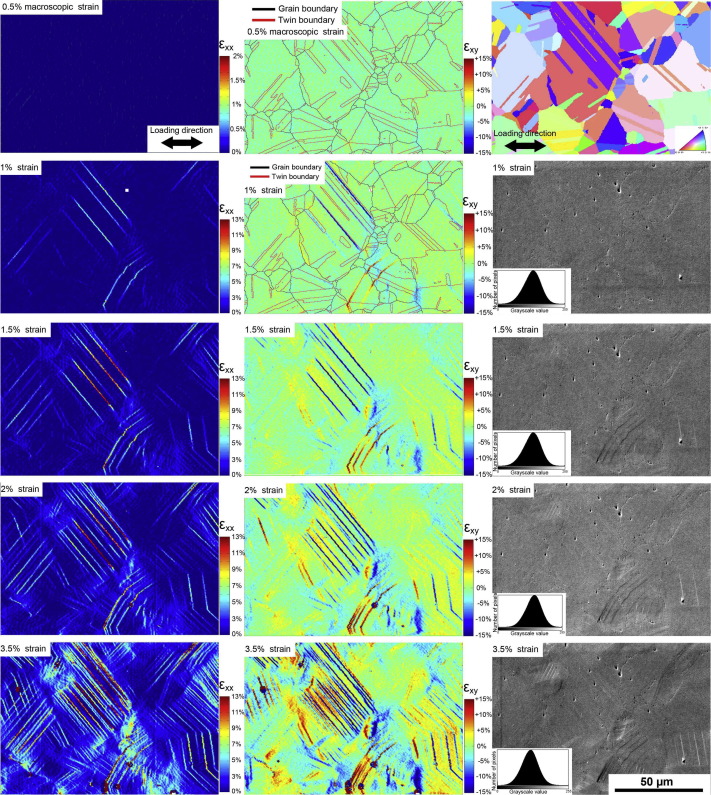}
	\caption{Tensile ($x$ direction represents the loading direction) strain field $\epsilon_{xx}$ (left), and shear strain field $\epsilon_{xy}$ (center) from DIC with images at high magnification after 0.5\%, 1\%, 1.5\%, 2\% and 3.5\% macroscopic strain in tension. Associated EBSD map (before loading) and SEM images at the different macroscopic strains are reported on the right. Reprinted from Ref.~\cite{stinville2015high}, Copyright (2015), with permission from Elsevier.}\label{fig:Bridier}
\end{center}
\end{figure}

%
%
\paragraph{Local field measurements through Digital Image Correlation:} The development of displacement field measurements has been expanding with the automation of quantitative image analysis techniques~\cite{Bornert2000, Pippan2011,Celotto2012, vanderesse2013open, Tasan2014, stinville2015high} such as Digital Image Correlation (DIC). DIC allows to go beyond the use of a regular grid and avoids some bias due to microgrids regularity and the fraction of surface covered by the grid lines. Speckle techniques are now regularly used to improve the accuracy of strain measurements at the grain level and at finer scales where plastic strain localization is manifested, for example, as physical slip bands~\cite{Zaafarani06ActaMat, stinville2015high}. Stinville \emph{et al.}~\cite{stinville2015high} studied damage accumulation processes in nickel-based superalloy using a DIC approach with speckle patterns achieved by etching. By combining DIC and EBSD (see Fig.~\ref{fig:Bridier}) they were able to detect low levels of strain and strain heterogeneities at the microstructural features. 

%
%
\paragraph{The popularization of EBSD measurements:} As seen in the examples above, microgrid and DIC techniques are generally combined with EBSD measurements. The development of EBSD in the last decade has allowed displacement field measurements to be complemented with crystallographic orientation field measurements~\cite{Adams1993}.
The automation processes and the reduction of probe size have enabled the mapping of fairly large surface area containing many grains with a spatial resolution now in the order of 0.1 $\mu$m~\cite{Zaefferer2003}. EBSD measurements recorded during the deformation lead not only to the knowledge of the crystallographic texture evolution, but also to information about heterogeneities~\cite{Zaefferer2003, Rollett2009, Pippan2011, Vignal2011, Bacroix2013, Jimenez2015}. As described later on in Section~\ref{sec:addressed}, such information is used to reconstruct ``synthetic'' equivalent representative microstructures.
Average misorientation metrics, such as the Kernel Average Misorientation (representing the numerical misorientation average of a given pixel with its six neighbors within an EBSD image), have been developed to quantify the correlation between plastic deformation and EBSD data. Orientation gradients and geometrically necessary dislocation (GND) densities can be calculated directly from the experimental measurements~\cite{Zaefferer2008, calcagnotto2010orientation, Wilkinson2010, ruggles2013estimations} and compared with non-local crystal plasticity formulations~\cite{liang2009gnd, Fressengeas2013}, for example.
Additionally, although not as popular yet, statistical analysis~\cite{bingham2010statistical,beyerlein2010statistical, juan2015statistical} (85 publications reported by ISI Web of Knowledge for the 2010--2015 time period in the research area solely limited to Materials Science on the topic of ``EBSD'' and ``statistical analysis'') on large EBSD datasets are being used to extract and study quantitative relationships between material microstructural features and the mechanical properties.
One of the main limitations with EBSD measurements lies in the degradation of the signal for highly deformed materials. Pixel indexing becomes difficult and the accuracy of the measurement is limited. Nevertheless, it is possible to overcome this difficulty by using the pattern quality index, which is another way to investigate the microstructure. Such a technique has been applied, for example, to recrystallization problems~\cite{Bacroix2005}: recrystallized grains are easily indexable while the non-recrystallized domains appear with a very low quality index.

%
%
\paragraph{Local field measurements through TEM-based approaches:} For very fine microstructures and/or local measurements, Automated Crystallographic Orientation Mapping obtained via TEM (ACOM-TEM) is an efficient tool~\cite{Zaefferer2000, Rauch2014}. Well-adapted for severely deformed samples (by equal channel angular extrusion, for instance) and for nanograin-size materials, this technique, similar to EBSD, allows for the collection of specific information on the mechanisms at the grain level and to confirm the model prediction at the same scale.
Finer spatial resolution of local misorientation down to a few nanometers is now accessible through the use of Precession Electron Diffraction (PED-ACOM). For example, Taheri \emph{et al.}~\cite{Taheri2015} accurately estimated dislocation density in a pure oxygen-free high thermal conductivity copper deformed by rolling and annealed at different temperatures from PED data using the Nye tensor. In addition to the popularity of these automated crystallographic orientation mapping techniques, other TEM-based approaches also enable accurate measurements of local fields a very fine scales. For example, HRTEM measurements of atomic plane displacements have been used to explore ferroelectric--ferroelastic interactions in PbTiO$_3$~\cite{Hytch1998}. The development of dark-field electron holography is another example of a technique used to map the strain field at the nanoscale~\cite{Hytch2014}. A recent publication~\cite{Beche2013} reviews the limitations and advantages of various experimental methods for measuring strain at this scale. The emergence of such techniques opens up new opportunities within the modeling community to validate and verify models.

%
%
  \subsection{3-D measurements and in-volume characterization}\label{sec:3Dmeasurements}
Surface observations, even during \textit{in situ} experiments, have proven inadequate to quantify microstructure evolutions and mechanical field measurements, especially for subsurface behavior. Planar sections through a volume can reveal information about the microstructural state, but local mechanical information can only be deduced through numerical simulations. Several options are available as alternatives to the displacement field measurements techniques described in the previous section. The advent of such techniques proved valuable within the modeling community.

\begin{figure}[!htb]
\begin{center}
	\includegraphics[width=8cm]{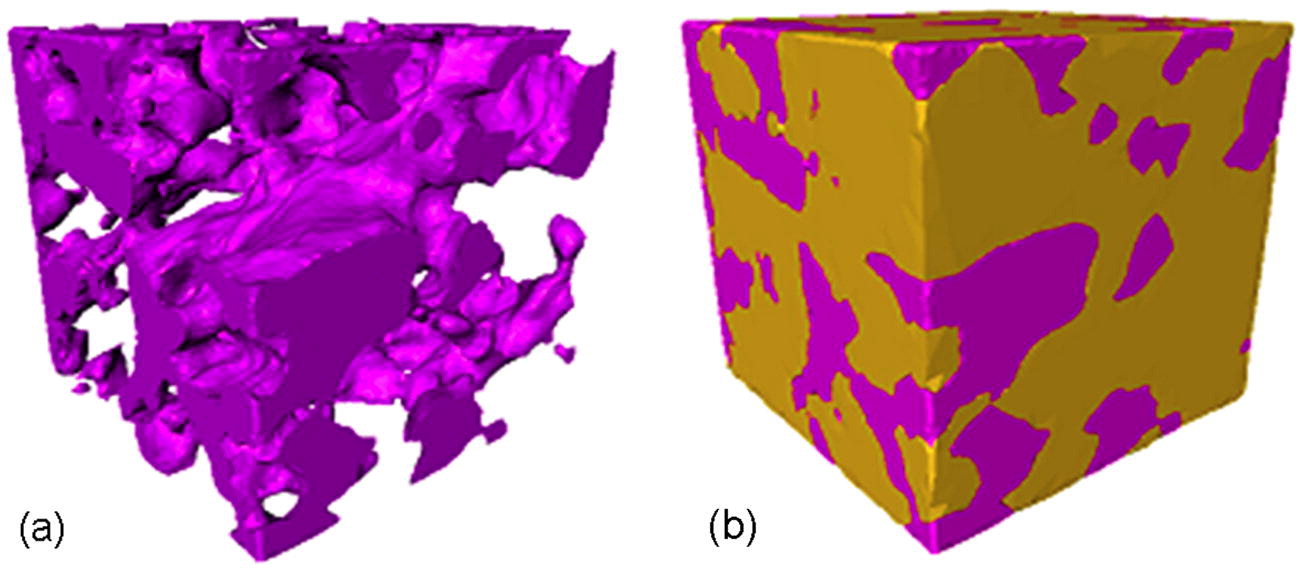}
	\includegraphics[width=8cm]{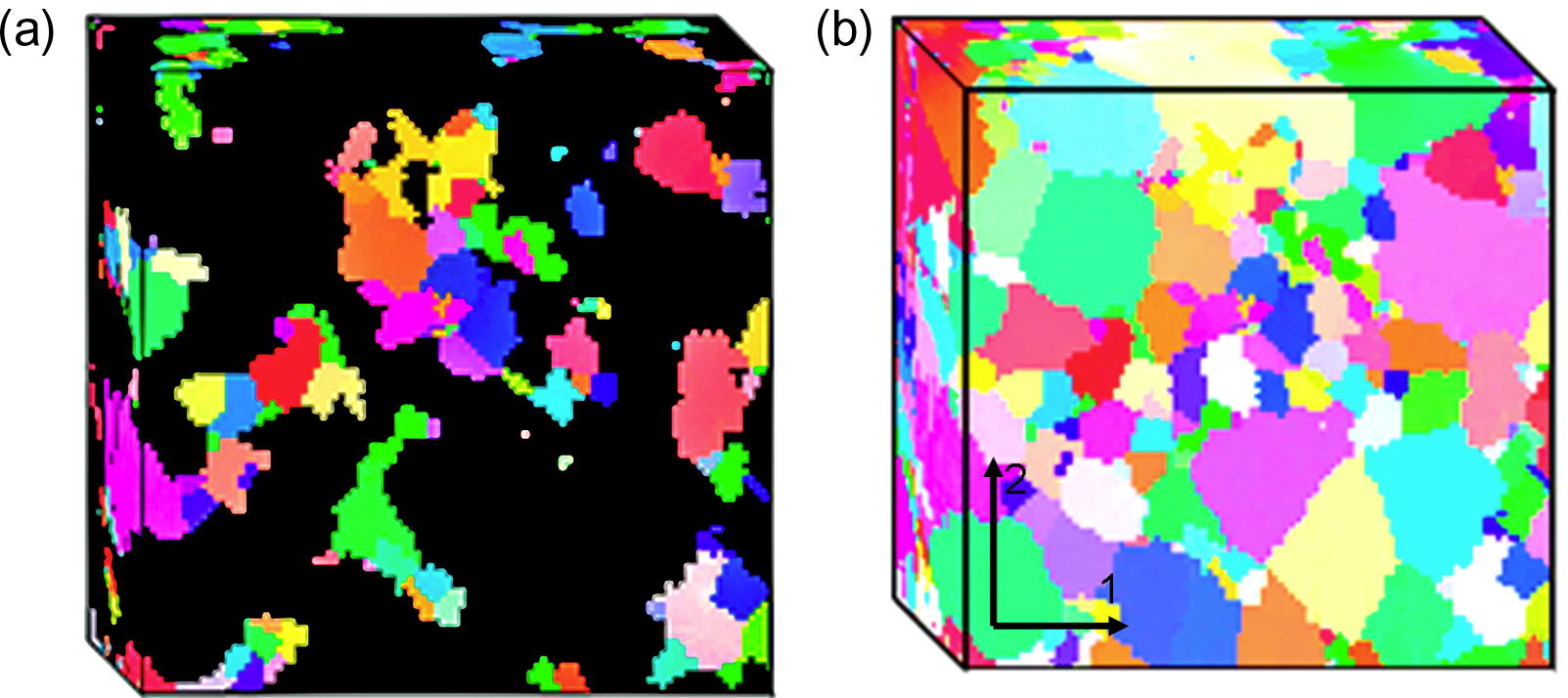}
	\caption{3-D phase distribution in a Mo-TiC composite. The top micrographs represent X-ray tomography of the TiC phase distribution (a), and the Mo-TiC composite (b). The specimen size is 30$\times$30$\times$30 $\mu$m$^{3}$. The bottom figures are composite reconstruction of 3-D sequential EBSD maps using a FIB for successive sectioning: (a) TiC phase, (b) Mo-TiC composite. The analyzed volume is 30$\times$30$\times$6.6 $\mu$m$^{3}$.Reprinted from~\cite{Cedat2012}, Copyright (2012), with permission from Elsevier.}\label{fig:3-D}
\end{center}
\end{figure}

%
%
\paragraph{3-D reconstruction:} Serial sectioning, often associated with EBSD measurements, allows for the reconstruction of the crystalline structure of a Representative Volume Element (RVE) and the associated grain orientations. This type of technique is widely used in conjunction with CPFEM simulations~\cite{Rowenhorst2006, Thebault2008, Cedat2012}. Recently, the fastidious tasks of repetitively polishing, drift correction and SEM observations have been replaced by \emph{in situ} focused-ion beam (FIB) layer removal combined with EBSD measurements~\cite{Uchic2006, Zaafarani06ActaMat, endo2014three, yamasaki20153d}. For example, microstructure reconstructions permit a way to evaluate, in the case of multiphase materials, the phase connectivity and associated influence on the macroscopic behavior and damage/fracture processes (Fig.~\ref{fig:3-D})~\cite{Cedat2012}.

While the experimental approach is well established, the 3-D reconstruction and the fine-scale definition of grain and phase interfaces are still difficult. In fact, as the phase and crystal orientation is recorded pixel by pixel, the resulting interfaces are irregular and composed of small flat rectangular facets~\cite{Rollett2009}. This is particularly detrimental for the accuracy of numerical calculations for phenomena addressing grain and/or interphase boundary effects on the stress-displacement field. The regular orientation of the facets with respect to the main macroscopic directions and the existence of sharp edges impact strongly the local estimation of the mechanical fields. This is of importance for simulation of any process in which interfaces play a major role (e\@.g\@.~intergranular fatigue damage, grain boundary diffusion, nanostructured materials). Efforts are presently dedicated to smooth the surfaces obtained through 3-D reconstruction techniques with numerical approaches, such as the level set technique, to overcome defects of both experimental digitalization and discrete numerical meshing~\cite{Bernacki2010}.

\begin{figure}[!tb]
\begin{center}
	\includegraphics[width=8cm]{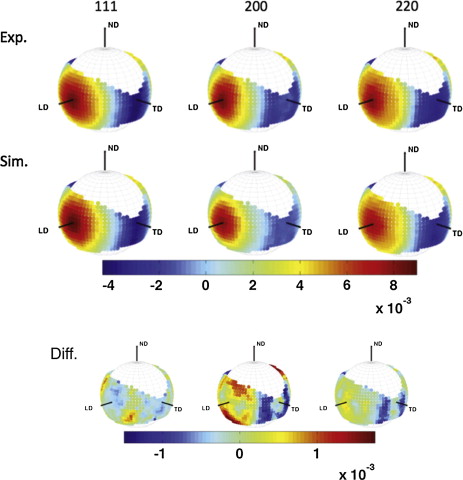}
	\caption{High energy X-rays combined with CPFEM\@: (Top) Lattice strain pole figures from experiment and simulation. (bottom) The difference between the experiment and simulation strain pole figures. Reprinted from Ref.~\cite{miller2014understanding}, Copyright (2014), with permission from Elsevier.}\label{fig:xrayCPFEM}
\end{center}
\end{figure}

%
%
\paragraph{In-volume characterization:} XRD and X-ray computed tomography provide allow for in-volume characterization~\cite{Buffiere2006, Evrard2010, oddershede2012measuring, Withers12, poulsen2012introduction, chow2014measurement, miller2014understanding, Pokharel14, olsson2015strain, Rollett2015}. The idea is fairly simple: X-rays are used to penetrate bulk metallic samples and the local stress/strain state is deduced from the evolution of the peak location and its enlargement---revealing the evolution of the microstructure.
A virtual representation of the microstructure is constructed and modeled using a model (most commonly CPFEM) to simulate the evolution of the internal state variable mediating the polycrystal. Simulations are then compared directly to experimental diffraction data.
For example, Oddershede and coworkers~\cite{oddershede2012measuring} measured the stress field around a notch in a coarse grained material using XRD and compared their measurements with a continuum elastic-plastic finite element model. The measured and simulated stress contours were shown to be in good agreement except for high applied loads, corresponding to experimentally observed stress relaxation at the notch tip. As illustrated in Fig.~\ref{fig:xrayCPFEM}, lattice strain pole figures (SPF) combined with CPFEM-based methodology~\cite{miller2014understanding} are also used for quantifying residual stress fields within processed polycrystalline components.

In the past, X-ray tomography has been limited to thin samples with a low density of defects, however with the development of near-field High-Energy Diffraction Microscopy (nf-HEDM) and the increasing availability of synchrotron sources, measurements of volumes on the order of a few $\textrm{cm}^3$ are now possible.  For example, this technique has recently been used to experimentally track the evolution of internal damage~\cite{Maire2014} or as a means to provide microstructural input (grain structure and crack morphology) for a CPFEM model embedded in a coarse homogeneous FE model~\cite{Spear14}. As pointed out by Whithers and Preuss~\cite{Withers12}, such experimental techniques, especially when combined with mesoscale models (e\@.g\@. CPFEM), are becoming increasingly popular for characterizing local phenomena such as fatigue crack growth or damage evolution. X-ray tomographic techniques are also used to obtain a complete 3-D characterization of architectured materials and track their evolution during deformation~\cite{Bienvenu2012}. Under certain conditions, \textit{in situ} measurements can be made, allowing for direct correlation with the numerical simulation results. XRD experiments combined with modeling capabilities offer a promising means to quantify mechanical fields but are so far limited by the time required to set up and run such experiments.
%
%
  \subsection{Toward 4-D characterization}\label{sec:4Dmeasurements}
\begin{figure*}[!tb]
\begin{center}
	\includegraphics[width=\textwidth]{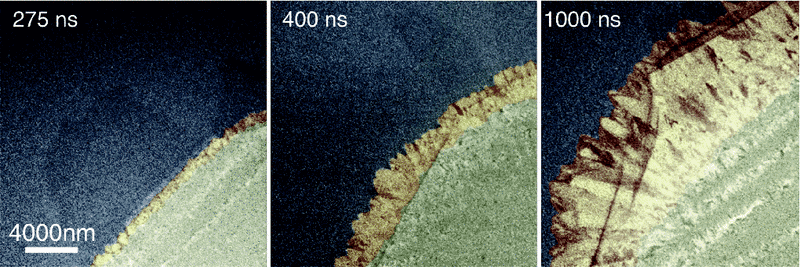}
	\caption{Dynamic TEM images of the time-evolving showing the growth dynamics of radial dendritic grains in germanium. Reprinted with permission from Ref.~\cite{lagrange2012approaches}. Copyright (2012), with permission from Elsevier.}\label{fig:PINEM}
\end{center}
\end{figure*}

Recent advances now not only enable 3-D measurements (see Section~\ref{sec:3Dmeasurements}) but they also integrate a fourth dimension---time---and track temporal evolution at resolutions as brief as picoseconds. Such characterization has been coined ``four dimensional (4-D) Materials Science.''
Notably, techniques such as hybrid environmental scanning electron microscopy (ESEM)-STEM tomography~\cite{jornsanoh2011electron, masenelli2014wet} and dynamic TEM~\cite{browning2012recent, flannigan20124d, baum2014towards} are emerging as high time resolution electron microscopy techniques with different variants enabling the real-time visualization of microstructure evolution.
It is anticipated that atom-probe tomography will gain this capability in the future~\cite{kelly2004first, isheim2006atom, kelly2007atom, Seidman2007,miller2012future}. A demonstration of the ultrafast temporal resolution achieved so far is shown for example in Fig.~\ref{fig:PINEM}~\cite{lagrange2012approaches}. The balance between space and time accuracies are well illustrated by \textit{in situ} X-ray tomography evolutions~\cite{Maire2014}. This allows \textit{in situ} recording of solidification, recrystallization, and phase transformation kinetics in polycrystalline materials.

The new opportunities offered by the experimental techniques and examples described in Sections~\ref{sec:atomiclevel}--\ref{sec:3Dmeasurements} illustrate how synergies between experiments and modeling can facilitate microstructure analysis (all the way down to the nanoscale). Motivated by a constant need to characterize deformation mechanisms more precisely, some improvements and iterations on the accuracy of both the experimental methodologies and corresponding simulations are necessary to evaluate the effects of microstructural variation on material response critically. The non-exhaustive list of 3-D and 4-D characterization techniques described above will require major scientific advances and improved resolutions (both spatial and temporal); from new modeling and experimental approaches to new instrumentation. However, as we will discuss in Section~\ref{sec:openquestion}, given the volume of data that can be acquired, such progress will need to be accompanied by (big) data management in order to efficiently and effectively collect, format, analyze, store, mine, and correlate large data sets linking process, microstructure, and materials properties~\cite{kalidindi2015materials}.
\section{Recent advances in modeling}
\label{sec:models}
The previous section highlighted recent experimental developments and illustrated how such measurements were integrated within corresponding models.
In this section, we review examples of modeling and associated numerical techniques using available experimental information across many length scales including (i) modeling addressing the effect of heterogeneous materials(Section~\ref{sec:addressed}); (ii) the development of approaches to model nanostructured devices and hybrid materials (Section~\ref{sec:multiphysics}); and (iii) the emergence of experimental measurement techniques incorporated into numerical frameworks directly~(Section~\ref{sec:virtualexperiments}).
%
%
\subsection{Modeling of heterogeneous materials}
\label{sec:addressed}
\begin{figure}[t]
\begin{center}
	\includegraphics[width=6cm]{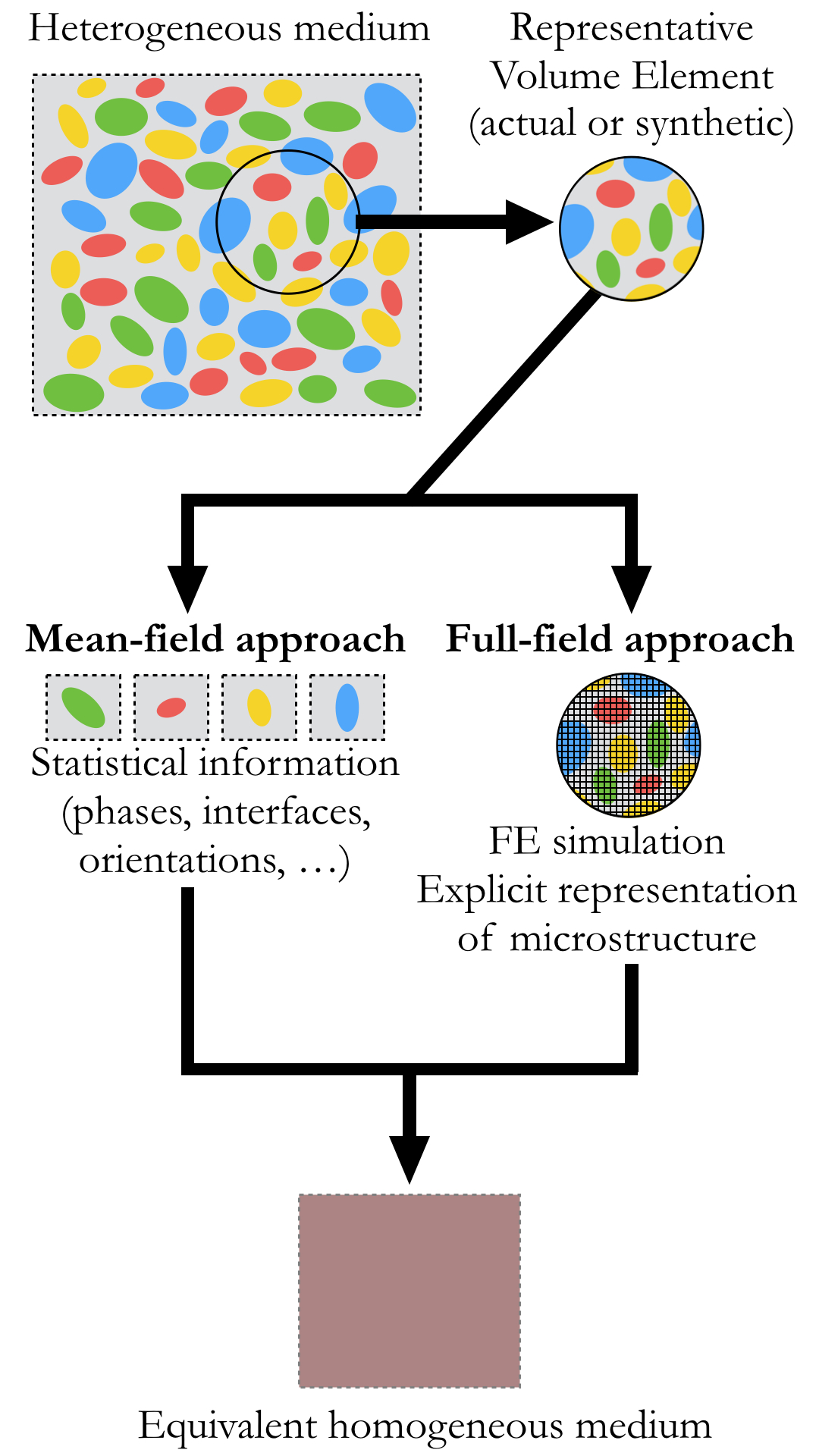}
	\caption{Mean-field versus full-field approach.}\label{fig:Mean_vs_Full}
\end{center}
\end{figure}
  
Understanding links between micro- and macro- scale properties is the key to predicting the behavior of media with a heterogeneous microstructure. As illustrated in Fig.~\ref{fig:Mean_vs_Full}, two general approaches using microstructural information exist in the literature:

\paragraph{Mean-field approaches,} initiated by Eshelby~\cite{Eshelby57}, are based on the solution of the inclusion of an ellipsoid (representing the grain) in the equivalent medium (representing the heterogeneous material, such as a polycrystal). They have been improved by, for example, taking into account elastoviscoplastic strains~\cite{Molinari97}, or the influence of various microstructure parameters such as the presence of different phases~\cite{Lebensohn97} and the grain size distribution~\cite{Berbenni07}. The main advantage of such approaches is their computational cost-effectiveness allowing microstructures of more than 1,000 crystalline orientations to be randomly generated and tested numerically; however, since the strain is assumed constant within each grain, this technique cannot account for intragranular heterogeneities and fine descriptions of the microstructure, such as real grain shapes and orientations.

\paragraph{Full-field approaches} have been developed to better take into account the latter specificities, by introducing relevant RVEs~\cite{Zeghadi07-2}. The simulation costs, however, severely limit the size of the simulation cell. Moreover, the question of the size of the RVE and its numerical homogenization remain open problems for polycrystalline materials: several multiscale strategies focusing on different numerical schemes can be used, e\@.g\@., multilevel FEMs, for which Schroder~\cite{Schroder14} gives a recent review, or so-called ``equation-free'' approaches, for which the Heterogeneous Multiscale Method (HMM)~\cite{E07} is well-known.

\paragraph{Microstructural representations} are required in either approach and are obtained either by using microstructures obtained from image-based experimental characterization techniques (see Section~\ref{sec:atomiclevel} on using TEM images, Section~\ref{sec:3Dmeasurements} on using X-ray-based measurements, and Section~\ref{modelinput} on using experiments as model input in general) or by generating a ``synthetic'' statistically representative microstructure numerically with varying degrees of randomness~\cite{Cailletaud03,Li10,Guilleminot11}.

\begin{figure}[!tb]
\begin{center}
	\includegraphics[width=8cm]{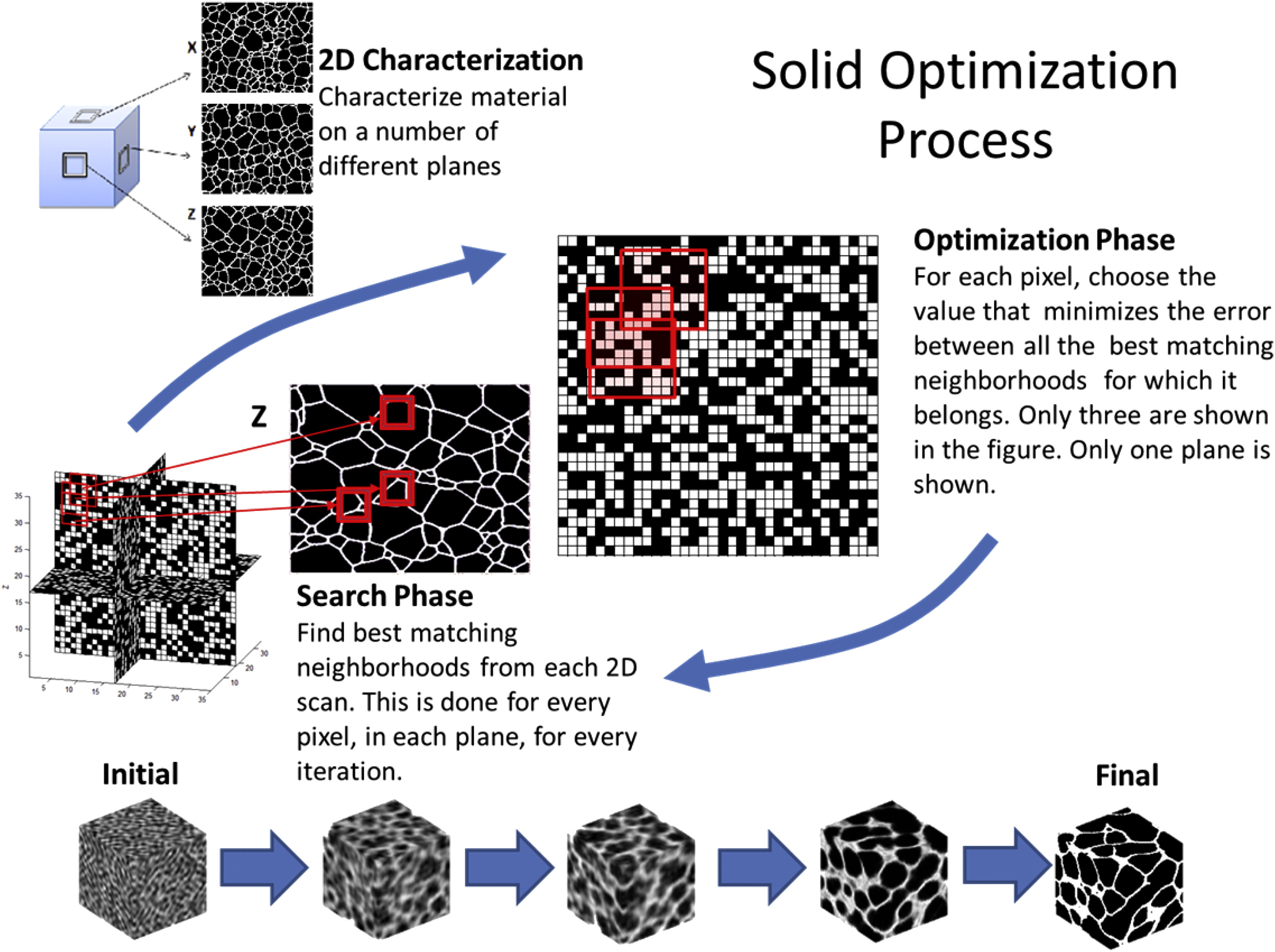}
	\caption{A schematic description of overall process for microstructure reconstruction using partial set of descriptors derived from 2-D characterizations. Reprinted from Ref.~\cite{turner2016statistical}, Copyright (2016), with permission from Elsevier.}\label{fig:microreconstruct}
\end{center}
\end{figure}

While appealing, exact numerical replicates of actual microstructures in models are nonetheless cumbersome and necessitate appropriate algorithms to process such big data. Many models use relevant epresentative volume elements (RVEs) as an alternative to determine the global macroscopic behavior of a heterogeneous material. The general approach is to use a small set of microstructural descriptors (e\@.g\@. determined from 2-D experimental cross-sections) covering features including material composition, dispersion, and microstructure morphology.  From this, statistically equivalent 3-D microstructures are then generated~\cite{fullwood2008microstructure, xu2014descriptor, turner2016statistical}.
As illustrated in Fig.~\ref{fig:microreconstruct}, a good example is the recent microstructure reconstruction algorithms based on a partial set of spatial correlations inspired by the field of solid texture synthesis~\cite{turner2016statistical}.
Recent works stemming from the modeling and applied mathematics communities are oriented towards numerical methods allowing for the automated conversion of experimental data to entities usable by FE simulations. For example, an anisotropic tessellation technique~\cite{Altendorf14} improved the reconstructions of actual microstructures from EBSD images, whereas an adaptive orientation reconstruction algorithm to get 3-D volumes~\cite{Li13} has been proposed for nf-HEDM images. Similarly, Lieberman~\emph{et al.}~\cite{Lieberman15} suggested an improved technique for determining the normal vectors to grain boundaries extracted from nf-HEDM acquisitions, allowing a much finer prediction of void nucleation sites.

It should be noted however that, while synthetic RVEs based on experimental statistical data have been considered historically (e\@.g\@. Ref.~\cite{HerreraSolaz14}), the limited size of such microstructures as being representative of the overall behavior may lead to outliers and often prove to be insufficient to obtain accurate local field information~\cite{Pokharel14}.

\paragraph{Homogenization and effective behavior of heterogeneous materials:} Many theoretical results exist as far as the homogenization of random media concerned: starting with linear elliptic equations, these results predict the existence of a constant deterministic tensor called ``effective tensor'', that yields a displacement field close to the solution of the original problem~\cite{Sab92,Bourgeat94,Cafferelli05}. However, the actual computation of the value of this effective tensor remains an issue, since its approximation converges more slowly as the size of the domain is increased~\cite{Bourgeat04}. Some recent alternatives exist in order to circumvent this drawback~\cite{Cottereau13}. It is also possible to go further and address a stochastic equivalent homogeneous model~\cite{Koutsourelakis07,Soize08,Clement13}, even if this has not been specifically applied to polycrystalline materials modeling yet.

There have been some attempts in the literature to truly couple probabilistic models, such as the multiscale stochastic FEM by Xu~\emph{et al.}~\cite{Xu07}. Other strategies are based on specific coupling schemes: whereas a surface coupling~\cite{Chevreuil13} was introduced to deal with localized uncertainties, specific applications of a superposition method, called the Arlequin method~\cite{BenDhia98}, have been proposed recently~\cite{Cottereau11, LeGuennec14}.

In addition to these multi spatial scale problems, it is also possible to address the simulation of physical phenomena exhibiting different well-separated time scales, such as damage evolution for fatigue loading. For the latter case, the adaptation to time of classical schemes of space homogenization is straightforward and leads to strong reductions in computational costs when compared with the initial full-scale problem: for example, Puel and Aubry~\cite{Puel14} give a tentative review of the so-called periodic time homogenization method. It is possible to deal with several scales simultaneously in time and space, e\@.g\@. Ref.~\cite{Fish12}. Similar adaptations can be proposed in the stochastic framework: for example, HMM can be used in MD~\cite{Ren07} or in kinetic Monte Carlo schemes~\cite{E07-2}.
%
%
\subsection{Mechanical modeling of nanoscale systems}\label{sec:multiphysics}
%
%
Modeling and simulations strategies described in Section~\ref{sec:addressed} are often challenged when the volume or the scale at which they are performed are of the same order as the material microstructures they must consider. Three domains where this is the case are considered here.

\paragraph{Nanostructured materials:} The low-dimensionality of nanostructures is constrained enough that continuum concepts are not applicable. Simulation tools such as atomistic modeling become necessary to simulate the elastic and mechanical properties~\cite{Gupta2013}. Similar constraints exist for nanosystems, such as NEMS~\cite{Romig2003, Lee2006} or nano-layered systems used for opto-electronics~\cite{Shaw2000}. However coupling and comparison to corresponding experiments at this length-scale are challenging~\cite{Eberl2009} due the general discrepancy in loading rates and time scales experimentally measurable and the ones achievable numerically.

\paragraph{Architectured or hybrid materials:} This class of materials motivates a lot of research in gaining materials properties not achievable through ``monolithic'' materials~\cite{Ashby2010, Ashby2013}. As such, using hybrid structures to expand the property space results in the development of new numerical and experimental multi-physics approaches. However, there is a wide-range of technical challenges yet to be addressed. Among these is the problem of similarity of length scale associated the structural topology of such materials and the one associated with the physical phenomenon that such material is trying to address. This is exemplified by gradient materials~\cite{Embury2010} or by nanostructural devices fabricated by 3-D printing. In some cases, the part thickness is such that the material is locally a single- or a multi-crystal with its own anisotropy and properties variations from one location to another~\cite{Ramirez2011, Simonelli2014}. Similar concerns hold in nanostructured materials in which the typical size of the microstructure is comparable to the length scale of deformation mechanisms (such as the deformation and fracture of nanocrystalline materials~\cite{Kumar2003} and heavily deformed pearlite wires~\cite{Shimokawa2014}). Again, atomic calculations are generally used in these systems.

\begin{figure}[!htb]
\begin{center}
	\includegraphics[width=8cm]{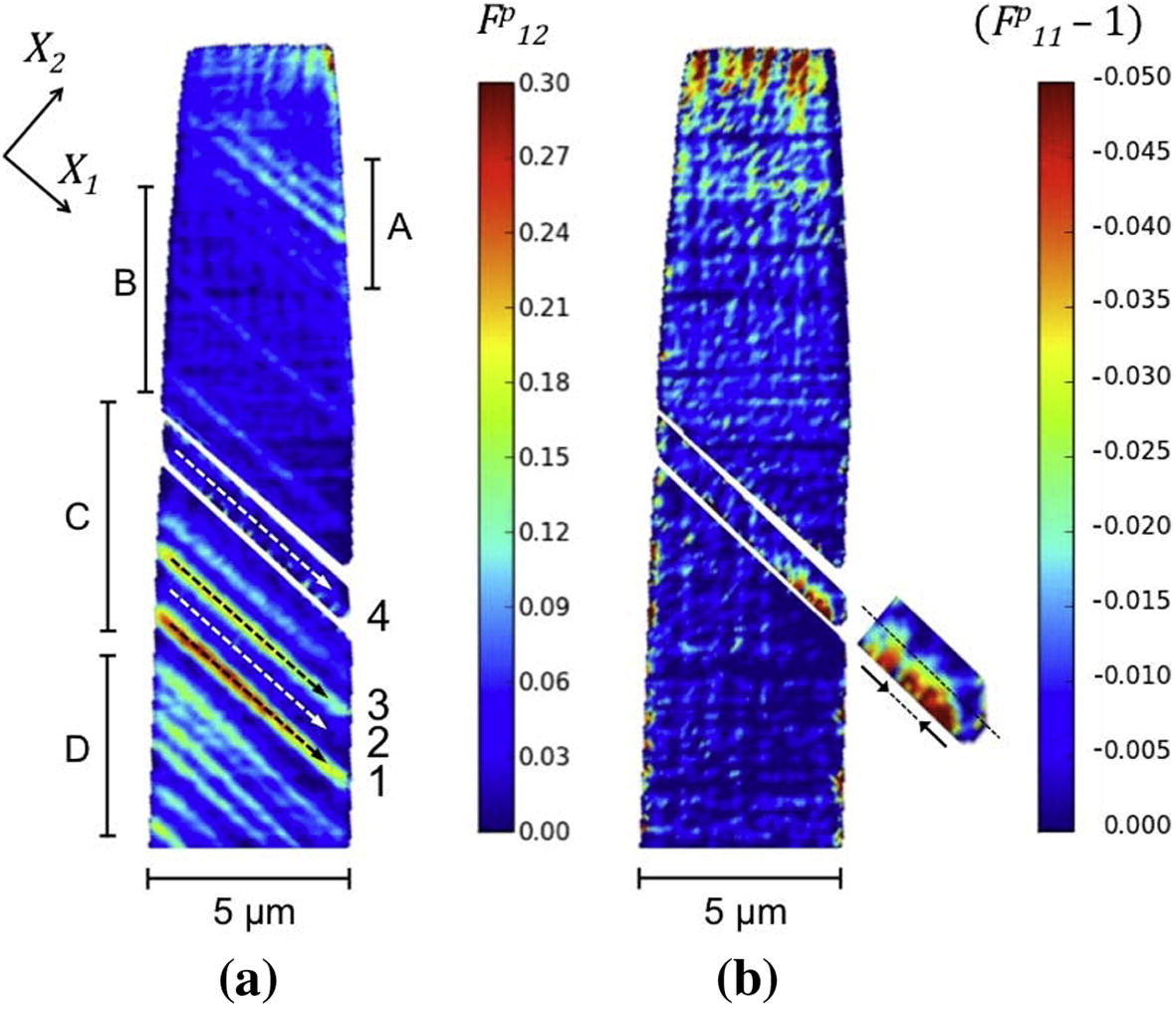}
	\caption{Map of localized slips along X$_1$ during compression of a copper micropillar. Reprinted from~\cite{Clegg2014}, Copyright (2014), with permission from Elsevier.}
	\label{fig:MicroPillar}
\end{center}
\end{figure}

\paragraph{Local investigation of the microstructure} are particularly important for nanoindentation measurements (discussed in Section~\ref{sec:atomiclevel}), where the atomic structure of the material has to be accounted for~\cite{Gouldstone2007}. Thus, MD or DDD calculations are used to estimate the material behavior below the indenter~\cite{VanVliet2003} or during micro pillar compression~\cite{Clegg2014}. Figure~\ref{fig:MicroPillar} shows a typical example of an experimental quantification of the displacement field in a copper micropillar; the measurement is deduced from DIC analysis of a speckle pattern deposited by FIB~\cite{Clegg2014}. Combinations of EBSD and displacement maps allow the local properties to be deduced by inverse calculation. Similar trends are noted for the analysis of HRTEM observations. A complete simulation at the atom and dislocation level is needed to quantify experimental observations~\cite{Pollock2013}. The main difficulties are then to define a representative volume that is large enough to include the local variation of the material, but small enough to limit the calculation size and time. Parallel calculations and model reduction are developed to enhance the simulation capacity. Nevertheless, the amount of deformation is still rather limited. The definition of the boundary conditions is very challenging and  this problem is so complicated that a multiscale strategy is not usually possible at present.
\subsection{Simulated images as a complement to experimental characterization}
\label{sec:virtualexperiments}
\begin{figure}[!htb]
\begin{center}
	\includegraphics[width=8cm]{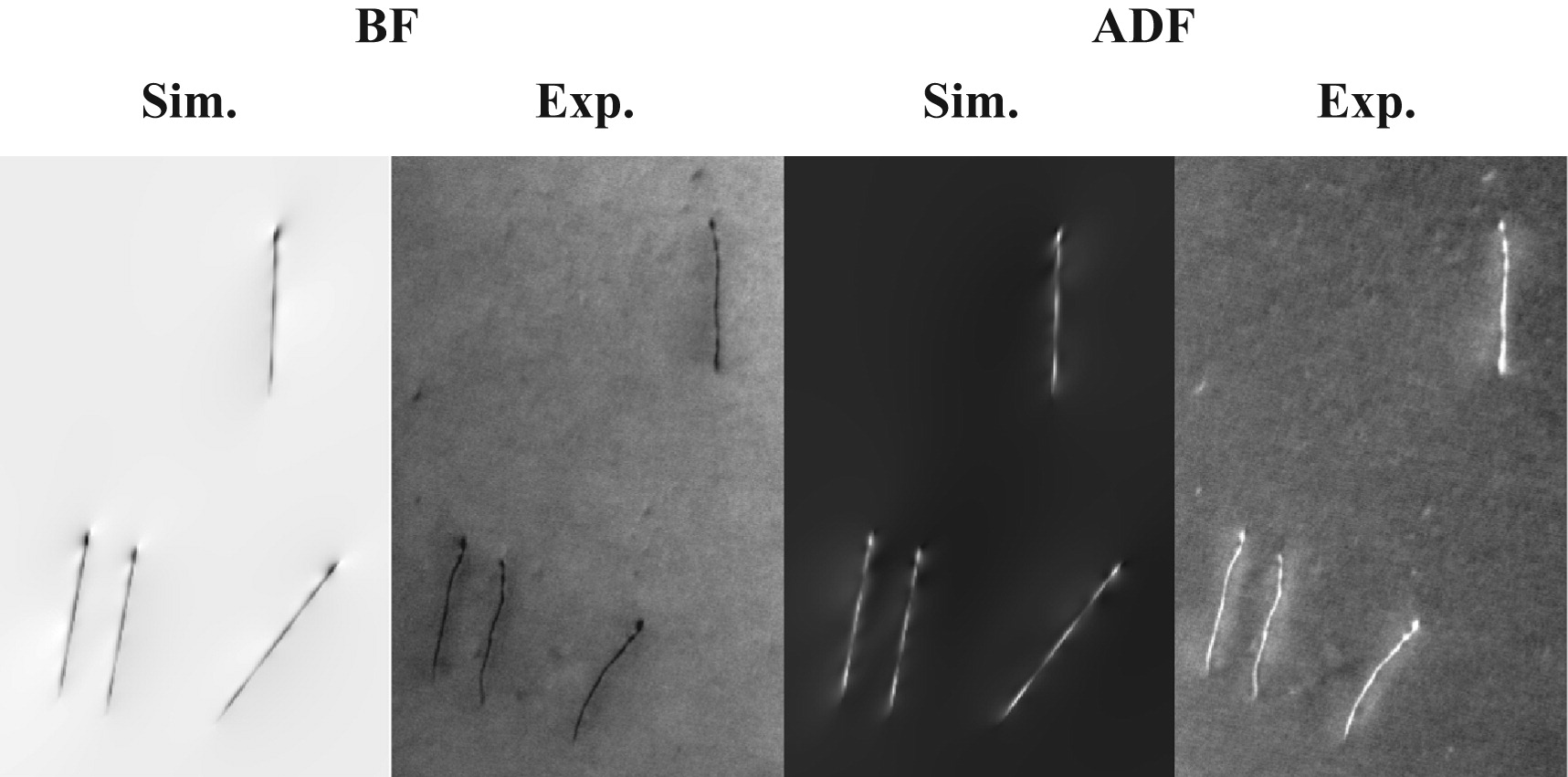}
	\includegraphics[width=8cm]{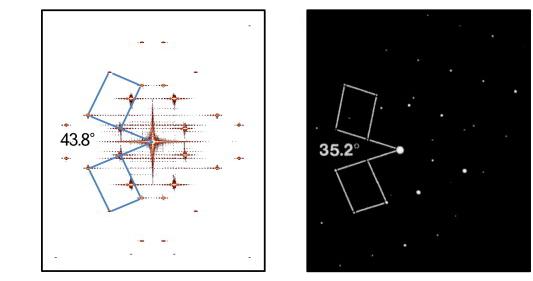}
	\caption{(Top) Diffraction contrast STEM of dislocations: imaging and simulations.  Reprinted from~\cite{phillips2011diffraction}, Copyright (2011), with permission from Elsevier. (Bottom) Comparison between virtual and experimental SAED patterns produced by (left) simulated and (right) experimentally observed in $\alpha$-Al$_2$O$_3$ $\Sigma11 (1~0~\bar{1}~1)(1~0~\bar{1}~\bar{1})$interface. Reprinted (adapted) from~\cite{coleman2015atomistic}, Copyright (2015), with permission from Elsevier.}\label{fig:virtual}
\end{center}
\end{figure}
Well-established experimental methods such as electron and X-ray diffraction are being implemented into computational models as a complement for the interpretation of experimental results. Simulated images can be done just like in real environments in which the model environment replicates the actual experimental settings and measurements acquisition. The goal of such modeling is two-fold: 
(i) verify the fidelity of the assumptions and results obtained by a given model and 
(ii) supplement experimental studies to reveal convoluted microstructural information collected by a given experiment.

%
%
Virtual diffraction~\cite{stukowski2009atomistic, coleman2015atomistic} patterns generated by atomistic simulation, is a good example of this. These models use classical algorithms from kinematic diffraction theory to create both selected-area electron diffraction (SAED) patterns and XRD line profiles within an atomistic simulations. While this idea has been around for quite some time in Materials Science~\cite{bristowe1980atomic, bristowe1984effect, oh1986structural}, the recent advances in virtual diffraction methods lie in the predictability of the methodology by performing virtual diffraction during an atomistic simulation via explicit evaluation of the structure factor equation without \emph{a priori} knowledge of the grain boundary structure (in contrast to legacy studies that computed the structure factor equation from known structures). Recent examples include studies of grain boundary structures~\cite{coleman2015atomistic} (see Fig.~\ref{fig:virtual}), microstructural evolution during deformation~\cite{higginbotham2013molecular, mogni2014molecular}, and deformation mechanisms~\cite{van2005situ, stukowski2009atomistic}.

%
%
Another great example of the replication of experiments within a computational framework is the STEM simulated image~\cite{leyssale2009image, phillips2011diffraction, robertson2013structural, Picard14} (see Fig.~\ref{fig:virtual}). Image simulations for the bright field and dark field TEM imaging uses a numerical algorithm based on the scattering matrix formalism within the context of the column approximation. For each image pixel, the algorithm computes a full convergent beam diffraction pattern. Such simulated images have been shown to be useful for crystalline defect analysis for a variety of diffraction configurations. Note that image simulation of TEM dislocation images has a long history dating back to the 1960s, however the recent novelty lies in the application to the particular mode of scanning TEM imaging rather than conventional TEM and again also in the predictability of the method rather than a mere confirmation of the experimental characterization.
\section{The role of experiments in multiscale models}
\label{sec:roles}
\begin{figure}[!htb]
\begin{center}
  \tikzstyle{place}=[ellipse,draw=blue!50,fill=blue!20,thick,minimum width=2.5cm]
  \tikzstyle{pre}=[bend angle=30,->,>=stealth,bend right]
  \tikzstyle{post}=[bend angle=30,->,>=stealth,bend left]
  \begin{tikzpicture}[auto,>=stealth]
    \node [place] (experiment) at (0:3) {\scriptsize\begin{tabular}{c} EXPERIMENTAL \\ MEASUREMENTS \end{tabular}};
    \node [place] (computer) at (120:3) {\scriptsize\begin{tabular}{c} COMPUTER \\ MODEL \end{tabular}};
    \node [place] (concept) at (240:3) {\scriptsize\begin{tabular}{c} CONCEPTUAL \\ MODEL\end{tabular}};

    \node (simulation) at (60:1.5) {\scriptsize\begin{tabular}{c} calibration, \\ model input \end{tabular}}
      edge(experiment)
      edge[->](computer);
    \node (programming) at (180:1.5) {\scriptsize\begin{tabular}{c} programming \end{tabular}}
      edge(concept)
      edge[->](computer);
    \node (discovery) at (300:1.5) {\scriptsize\begin{tabular}{c} model development \\ through discovery \end{tabular}}
      edge(experiment)
      edge[->](concept);
    \def\myshift#1{\large\raisebox{1ex}}
      \draw[<->,bend left, bend angle=60,postaction={decorate,decoration={text along path,text align=center,text={|\myshift|Validation}}}](computer) to (experiment);
      \draw[<->,bend left, bend angle=60,postaction={decorate,decoration={text along path,text align=center,text={|\myshift|Verification}}}](concept) to (computer);
      \draw[<->,bend left, bend angle=60,postaction={decorate,decoration={text along path,text align=center,text={|\myshift|Confirmation}}}](experiment) to (concept);
  \end{tikzpicture}
  \caption{Schematic of information flow between experiments, conceptual models, and computer models. After Ref.~\cite{Schlesinger1979}.}\label{fig:roles}
\end{center}
\end{figure}
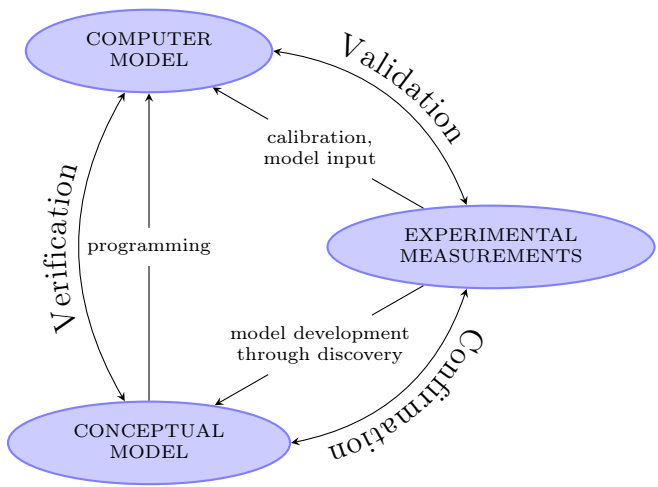
Figure~\ref{fig:roles} highlights the information flow between experiments and models.  The National Research Council recognized three distinct purposes for experiments in ICME efforts: 
(i) classic validation showing whether the models are consistent with what can be measured; 
(ii) model input that can set, e\@.g\@., model scaling parameters or microstructure; and 
(iii) collection of large amounts of data that can be mined~\cite{CICME2008}.
This section explores each of these in turn, but McDowell and Olson have noted that ``a combination of expert estimation based on prior experience, experiments, and models and simulations at virtually
every mapping'' is used in practice~\cite{McDowell-2008}. In many cases, these three modes of using experimental data are used concurrently and iteratively to improve the accuracy of the model and to bound our certainty in its predictions.
%
%
\subsection{Calibration, verification and validation}
Different communities use verification and validation\ (V\&V) to accomplish different things~\cite{Thacker-2004, Oberkampf-2004}.
While software developers use V\&V to ensure the reliability and robustness of their code, the modeling community uses the process to quantify and improve the agreement between model output and experimental measurements (often with the additional hope that improving the model to address questions that are measurable experimentally will help to make improvements in predictions to questions that are more difficult to assess).

In building new models and simulations, we start first with conceptual models describing the real world (i\@.e\@.~the experiments in the mind of the modeler). These conceptual models include the relevant partial differential equations, initial conditions, and boundary conditions. The conceptual model is implemented as a computer model to solve these complex systems efficiently. Verification is the process of determining that this computer model is an accurate solution to the conceptual model. However, this process is disconnected from the real world, so experiments provide no role in the verification process.
Observational experiments may precede the conceptual model by providing mechanism discovery (e\@.g\@.~in \emph{in situ} TEM) or be used to confirm the conceptual model is a good description of reality, but they do not assess whether a numerical implementation of a mathematical model is correct.

The computer implementation may have adjustable parameters (either numerical or physical) and calibration experiments may be used to set these. Only after a simulation tool is verified and, if needed, calibrated, is it validated.
Validation is the process that quantifies how accurately the computer model represents the real world. Rather than asking if we solve the equations correctly, it asks if we're solving the correct equations. Validation is typically done through a series of high quality experiments under the same conditions (when possible) imposed by the model and it should use different data than any calibration data. It is often possible to break models up hierarchically and to validate aspects of the model~\cite{Cowles-2012}.

The validation requirements for lower level aspects (e\@.g\@.~nanoscale) are generally more stringent than for the system level aspects (e\@.g\@.~continuum scale) because errors propagate up the hierarchy. In multiscale models, this often means very accurate measurements must be taken at very small length and time scales. Excellent validation has been performed using nanoindentation experiments, for example~\cite{Wei-2012,Pen-2011}. Unfortunately, validation at these short scales is not always easy. Available small length-scale models (e\@.g\@.~MD) tend to also operate over very short time periods and available longer length-scale models (e\@.g\@.~CPFEM) tend toward coarser time scales as well.  Conversely, experiments that have very high spatial resolution (e\@.g\@.~APT) tend not to be able to measure fast changes and current techniques to observe faster kinetic processes (e\@.g\@.~nanoindentation) tend to be limited towards longer length scales. Furthermore, whereas MD may describe behavior in the picosecond to nanosecond regime, our time-resolved measurements tend to be limited to the microsecond to millisecond regime~\cite{Gates-2005}.
%
%
\subsection{Model input}\label{modelinput}
The previous section touched on using calibration experiments to set adjustable parameters in a model. Experimental data may enter models for another reason. While multiscale models can often make predictions from first principles, predictions can become more precise or the calculations can be simpler if experimental input is used for scaling or to obtain material features that models don't predict well. Atomic scale models depend upon interatomic potentials, and these may be obtained from experimentally-determined lattice parameters, elastic constants, vacancy formation energy, and bonding energies. Potential fitting has been built into codes such as \textsc{gulp}~\cite{Fan2011} or \textsc{lammps}~\cite{plimpton1995fast}, for example.
At the mesoscale, fatigue and fracture are particularly sensitive to material defects. So, in fatigue models~\cite{McDowell2003,Horstemeyer2010}, for example, crack and pore size parameters have been set from experimental values.

Regarding the use of experimentally-obtained microstructural images as model input, there have been examples in the literature of using  2-D images for either 2-D mechanical property models~\cite{Chawla2005} or to simulate a 3-D microstructure that matches the statistical properties of experimental datasets~\cite{Roberts2001,Ghosh2004}. As discussed in more details in Section~\ref{sec:3Dmeasurements}, full 3-D microstructures are now being used in models. Destructive analysis using serial sectioning techniques has allowed for the more accurate and quicker prediction of elastic moduli~\cite{Chawla2004}, the onset of plasticity, and fracture~\cite{Lewis2006}. Non-destructive techniques, such as X-ray tomography have allowed identical microstructures to be used in both modeling and experimental determination of mechanical properties~\cite{Maire2003, Youssef2005, Kenesei2004}.

Combining 3-D and nanometer-scale analysis shows promise and there has been recent work that uses APT data in MD~\cite{Larson2013236, Prakash-2015} and DDD~\cite{Krug-2014}. The APT's imperfect detection efficiency (modern instruments detect roughly half of all atoms that are field-evaporated) and imperfect spatial resolution have motivated a hybrid approach that combines atom positions determined experimentally by APT with Monte Carlo simulations to create input suitable for density functional theory calculations~\cite{pareige2011kinetic,Moody-2014}.

%
%
\subsection{Experimental discovery and data mining}
Because sensors, automated data acquisition systems, and data storage have become faster and less expensive, modern experimental methods generate a huge amount of data. Experiments have led to new discoveries directly where new mechanisms and behavior are observed by a researcher who develops a conceptual model. But experiments now also serve as the basis for discovery through automated statistical analysis. The two challenges of data mining are first generating the data and then analyzing it efficiently.

%
%
\paragraph{Experimental data collection:} One method to generate a huge quantity of data is through a combinatorial study. Experiments can systematically vary parameters. Even a single parameter, varied manually, can lead to deep insights. By varying micropillar dimensions cut by FIB milling, for instance, the size limits for plastic processes can be found~\cite{Uchic2004}. But when high-throughput characterization techniques are used (that can often be conducted in parallel and/or be scripted to run in a serial fashion without interrupting the researcher), an even larger swath of variable space is explored~\cite{Broderick2008}.

%
%
\paragraph{Data reduction and analysis:} Sifting through this vast amount data poses a second challenge. One common data reduction technique is Principal Component Analysis (PCA)~\cite{Rajan2002,Broderick2008}. Multivariate data often has many correlated variables that are responsible for an observation. PCA transforms those variables to a new coordinate system where the new variables are linearly uncorrelated and sorted from greatest-to-least variance.
\section{Open scientific questions}
\label{sec:openquestion}
The study organized by The Minerals, Metals \& Materials Society (TMS) entitled \emph{Modeling Across Scales: A Roadmapping Study for Connecting Materials Models and Simulations Across Length and Time Scales}~\cite{TMSMultiscale} identifies 30 critical gaps and limitations for materials modeling across multiple length and time scales and include the inefficiencies in application programming interfaces, limitations in CPFEM analysis, the need for multiscale experiments to calibrate and validate multiscale models, or the coupling of models and experiments for performance predictions of rare events.

As discussed in the above sections, open questions lie in the different complementary domains detailed in this manuscript: characterizing initial microstructures and their temporal evolution experimentally, modeling coupled physical processes over time, linking different length and time scales, and managing large volumes of data. Without being exhaustive, a few examples of major challenges are raised.

%
%
\subsection{Gaps and limitations with current experimental instrumentation and tools}
\label{sec:investigation}

%
%
\paragraph{Measurements of stress fields within a heterogeneous medium:} Experimental visualizations of the displacement and rotation fields are now quite common and provide important microstructural information. However, they are still limited in the sense that deformation is not a state variable from a theoretical point of view and the deconvolution between the elastic and plastic components of the strain field is difficult. An associated challenge is the measurement of the local stress fields with high spatial and temporal resolution. These measurements have to be done under mechanical loading, i\@.e\@. during \emph{in situ} experiments. This is of particular importance for the stress concentration determination close to phase- or grain- boundaries for example; a research topic on experimental explorations closely linked to damage and fatigue behavior, for instance.

One of the main difficulties is to combine a fine spatial resolution (on the order of hundreds of nanometers or below) with fine time resolution measurements. Presently, conventional XRD allows an acceptable accuracy for the estimation of the stresses through the quantification of the atomic plane spacing (see paper by Chang \emph{et al.} published in this special journal section), but the collection time is still prohibitive. On the hand, synchrotron XRD, which allows 3-D local investigation~\cite{JuulJensen2008}, is not yet accurate enough to allow such a measure.

Closely related to these experimental roadblocks, the automation of data acquisition and analasis has to be developed to analyze many diffraction patterns in a relatively short time. Such improvement is critical because the required accuracy is very high and the signal profiles and maxima are not easy to define without any ambiguity. A huge volume of data will be generated from which we must extract the most important information (see Section~\ref{sec:data}). The improvement of such measurements will not only offer more experimental data on complex materials systems more easily and contribute to the V\&V of models, but it will also drive the development of more complex anisotropic models.

%
%
\paragraph{Improvement of the accuracy and sensitivity of non-destructive 3-D measurements:} 2-D investigations and destructive 3-D analysis examining 2-D sections may introduce a bias due to free surfaces where mechanical and environmental states are quite different from the bulk.
These effects can generally be accounted for in simulations, but the actual mechanisms may still be misunderstood. While computational power and algorithms are allowing 3-D simulations that yield greater insights, non-destructive 3-D characterizations must be made available with a similar accuracy and sensitivity. During the past ten years, many new developments have been made through synchrotron radiation. Some recent examples in the literature demonstrate local misorientation measurement~\cite{Rollett2015} and the localization of nano-cracks by X-ray tomography~\cite{Maire2014}.

Improvements can also be made in the speed of some experimental measurements, which tend to be much slower than desired strain rates. Ideally, the measurements would be coupled with \emph{in situ} deformation and could be done on the fly as the deformation occurs. This implies the development of mechanical experiments on small samples deformed with devices compatible with the diffraction and other investigation tools. The limitation will probably be defined by the representative volume which allows a continuous mechanical description (see Section~\ref{sec:mechanics}).

%
%
\paragraph{Combination of simultaneous measurements at the same location:} An additional challenge now is to compile different types of 3-D measurements of the same sample (ideally, simultaneously) to determine local microstructure--properties relations. 
There are now efforts to correlate APT data with electron tomography and FIB measurements to improve reconstructions, for example~\cite{Babinsky2014}.

A greater ability to combine information from multiple analysis techniques and a greater ability to perform the techniques simultaneously would be powerful additions to our toolbox. Ideally, one would be able to measure not only chemical segregation, but also correlate that with the knowledge of local strains (or elastic stresses), particularly close to phase or grain boundaries where segregation can play an important role for mechanical localization. Similarly, our models may benefit if grain character could be correlated with local yield strength.
%
%
\subsection{Gaps and limitations with current models}
\label{sec:mechanics}

%
%
\paragraph{Scale multiplicity due to microstructural evolution and materials response:} The core principle that materials properties are determined by composition and structure requires that predictive models are able to treat multiple scales within the same paradigm consistently. While many macroscopic behaviors result from the average of local variables, such as the effect of grain size or phase fraction on the stress--strain curve, some are directly linked to specific localizations that result from microstructure distribution and the chance of deformation, impurity diffusion, phase changes, etc. This dependence on localized quantities is the norm for damage, fracture, and fatigue. Because these may arise in a large structure, one has to combine a complete macroscopic description that accounts for boundary conditions with an accurate quantification that allows local calculations. For a nanosized sample, the characteristic volume of the microstructure is of the same order of the representation volume, but for meso- or macro- scopic samples, a complete description of the microstructure at a small scale is impossible. How can we overcome this problem? Periodic structure assumptions and/or homogenization are possible, but are not accurate for local phenomena such as damage, cracks, precipitates, or recrystallization nuclei. There is a real theoretical challenge to represent the whole microstructure with its average heterogeneity. Current models employ discretization methods on a specific grid size and time marching strategy. The major challenge associated with dealing with disparate scales within the same framework resides in the fact that different scales require different grid resolution and time steps, making the scale bridging a complicated endeavor. A spatial example includes first- and second- generation twins in alloys~\cite{juan2012prediction, juan2015statistical, abdolvand2015study} while a temporal example is the phenomena associated irradiation with ultrafast defect generation (displacement cascade) versus the slow defect migration and interaction~\cite{woo1992production}. Such considerations will undoubtedly need to be developed in conjunction with the development of new instrument while expanding their accuracy and sensitivity across scales.

%
%
\paragraph{Quantification and propagation of uncertainties:} Many fundamental processes and mechanisms, when viewed at the appropriate scale, behave stochastically. Atomic diffusion is one example and yet the collective behavior of large numbers of diffusing atoms can, on average, be quantified deterministically. Materials composition too can, depending on the scale, be considered highly heterogeneous. Uncertainty quantification and its propagation across scales (time and space) are both very challenging. Indeed, uncertainty at different scales needs to be considered differently. The increasing cross-pollinations between the modeling community and the experimental community will also need to bring in the statistics and probabilistic communities in order to address the variability and predictability of materials response over multiple scales. Integrated rigorous statistical paradigms will ultimately permit the identification of critical microstructural features that are vital to property evolution, especially at the system level, leading to materials design solutions.

%
%
\paragraph{Inefficiencies of current approaches for multiscale modeling:} A wide range of multiscale modeling approaches is limited by the increasing computational expense and cumbersome algorithms associated with dealing with multiple scales and/or large sets of experimental data that need to be incorporated into the models. This include pre-conditioners, i\@.e\@. reliable mathematical techniques conditioning the input data and problem into a more suitable form for numerical simulations or the improvement of reduced order modeling methods (i\@.e\@. reducing the degrees of freedom) enabling limitation of resolution losses. An additional motivator to address these inefficiencies is the promise that early model results may help to design a set of experiments that explores critical parts of the parameter space rapidly and inexpensively, leading to further model improvement.
%
%
\subsection{Gaps and limitations with data management}
\label{sec:data}
%
%
\paragraph{Storage and sharing of ``big data'':}  The volume of data (both from numerical models and experimental measurements) that will be acquired within this emerging field of Materials Science is undoubtedly huge. The heterogeneous, complex and convoluted nature of the data collected will need to be distributed across many communities (see Section~\ref{sec:intro}). Resolving the challenges associated with the data management will require the establishment of common standards and protocols to vet, compare and use such data reliably. Such mechanisms need to be put in place for the community to not only allow to compare, verify, and validate data used in models, but also to provide means to make decisions based on greater amounts of information entering the model (with no need for recalculating/charactering). Kalidindi and De Graef~\cite{kalidindi2015materials} provide a more detailed review on important aspects of materials data management, such as storage hardware, archiving strategies, and data access strategies. Marsden \emph{et al.}~\cite{Marsden-2014} highlight the breadth of data that must be stored for multiscale materials modeling and the challenge of making this data interoperable with many computational tools.

%
%
\paragraph{Speed up engineering transfers of fundamental approaches:} Even though Materials Science is contributing to knowledge development, its final goal is to provide technological innovations and practical tools and knowledge relevant to the industry. One of the main practical advantages of multiscale approaches is orienting new material developments through an inverse approach. That is, knowing the exact distribution of final properties expected in a part, to be able to design the optimum material, combining graded properties and multi-material solutions. Such an approach gives rise to guidelines for chemical composition and processes for new alloys and materials. However, it is untenable to implement complex multiscale, multi-physics codes in an industrial context. It is then of importance to develop simplified simulations integrating most of the specificity of the fundamental codes. A complementary method is to design coupled approaches for which different codes interact at different scales. For instance, a 2-D finite element code with macroscopic behavior laws may identify the most critical part during forming of a complex part, generates the boundary conditions to a specific volume, and a more complete 3-D physics-based model can probe the part. Some of these methods are already in place, but some work has to be done to improve interactions between the different scale models and to speed up calculations. Finally, databases have to be created and qualification of already existing materials is needed in order to reduce the domain that models would explore. 

%
%
\paragraph{Inefficiencies of programming interfaces:} Portability between different codes, algorithms, and instrument interfaces are not always available and/or are inefficient because of computational complexity to map data into a usable form. For example, phase field models coupled with thermodynamic databases for the multi-component systems lack the ability to interface efficiently and information transfer from one to the other is often cumbersome and amateur. The challenge associated with this interfacing problem is to devise bidirectional methods and protocols to efficiently communicate and pass data. This statement is true not only for interfaces between two models but also for models directly incorporated with experimental setups~\cite{Picard14}. 
\section{Concluding remarks}
Multiscale modeling approaches have proven to be powerful methods to predict macroscopic mechanical responses for a variety of loading conditions. They rely on experiments for validation, model input, and mechanism discovery. As these approaches are extended to new problems---new material systems, environments, and/or forms of mechanical response---new models, algorithms, and experimental techniques are needed and will also need to be integrated within the same framework in order to meet the increasing demand for better insights and predictability of the performance of such materials systems.

Recent developments in experimental instrumentation have helped build better multiscale models. Breakthroughs in materials research instrumentation now enable us to link atomistic, nanoscale, microscale, and mesoscale phenomena across time scales. Emerging research
opportunities for experimental improvements include increasing resolution and fidelity, decreasing measurement time, automation of analyses, and better management of the large amounts of data.
Such improvement will undoubtedly lead to a much deeper understanding of the local phenomena associated with heterogeneities and interfaces. Finally, 3-D investigations are now largely able to quantify microstructural evolutions in bulk and not only on the sample surface with the influence of free surfaces, or on a section made though destructive preparation.

Along with these experimental advances, numerical and computational improvements enable the study of complex materials systems across multiple length and time scales. Even though the available experimental and computational tools provide incomplete and uncertain information on phenomena of interest, improvements in the affordability of computational power and storage capabilities allow the simulation of large representative volumes capturing more and more microstructural features. New algorithm developments speed up the calculation and lead towards time scales of simulations that are compatible with experiments. One of the main challenges associated with these models is related to connecting information across scales to reduce the amount and complexity of the
information at higher scales: we have no \emph{a priori} knowledge of which physics and variables at the lower length scales are ``worth'' upscaling. As such, emerging research
opportunities for modeling improvements include novel approaches to identify this information.

Finally, both of these improvements are accompanied by an increasing focus on materials data analytics, which focuses on data-driven approaches in Materials Science to extract and curate materials knowledge from available data sets (from numerical and experimental studies). Opportunities in this field of research include the standardization and automation of workflows to handle ``big data'' and develop protocols and strategies to take advantage of emerging high performance data science toolkits and cyber-infrastructure

Despite the impressive progress made during the last decade, the number of materials features describing the links between processing, structure, and properties is too large to allow systematic study using established scientific approaches with modest resources. The Materials community is still facing huge challenges as legacy data, numerical models, and multi-physics knowledge are fragmented and compartmentalized across so many disciplines. Research at the crossroads between computational modeling and experimental characterization will continue to have a broad impact on a wide range of science and engineering problems in Materials Science, but only through an efficient and interactive interdisciplinary collaboration. Given this trend, the development of modern data science and cyber-infrastructure is critical to mediate and accelerate such collaborations.
\begin{acknowledgements}
This review article was written by the organizers of a symposium on the synergies between computational and experimental characterization across length scales at the 7th International Conference on Multiscale Materials Modeling, October 6--10, 2014 in Berkeley California USA\@. This symposium provided a forum for the Materials Science community to present and discuss the recent successes of predicting various physical phenomena and mechanisms in materials systems enabled by the collaboration between experimentalists and modelers. Some scientific research findings, successful collaborations, and tools leveraging the experiment-modeling synergy presented during this symposium are discussed in the present manuscript. Consequently, the authors thank all participants of this symposium for inspiration and motivation.

RD and RAK are supported by the Laboratory Directed Research and Development program at Sandia National Laboratories, a multi-program laboratory managed and operated by Sandia Corporation, a wholly owned subsidiary of Lockheed Martin Corporation, for the U.S. Department of Energy's National Nuclear Security Administration under contract DE--AC04--94AL85000.

The authors declare that they have no conflict of interest.
\end{acknowledgements}

\bibliography{\jobname}

\begin{thebibliography}{100}
\expandafter\ifx\csname urlstyle\endcsname\relax
  \providecommand{\doi}[1]{doi:\discretionary{}{}{}#1}\else
  \providecommand{\doi}{doi:\discretionary{}{}{}\begingroup
  \urlstyle{rm}\Url}\fi

\bibitem{Ohashi09IJP}
Ohashi T, Barabash RI, Pang JWL, Ice G, Barabash OM (2009) {X}-ray
  microdiffraction and strain gradient crystal plasticity studies of
  geometrically necessary dislocations near a {N}i bicrystal grain boundary,
  Int J Plasticity 25:920--941

\bibitem{Zaafarani06ActaMat}
Zaafarani N, Raabe D, Singh R, Roters F, Zaefferer S (2006) Three-dimensional
  investigation of the texture and microstructure below a nanoindent in a {Cu}
  single crystal using {3D EBSD} and crystal plasticity finite element
  simulations, Acta Mater 54:1863--1876

\bibitem{VonArdenne1938ZFP}
Von~Ardenne M (1938) Das {E}lektronen-{R}astermikroskop, Z Phys 109:553--572

\bibitem{Seidman2007}
Seidman DN (2007) Three-dimensional atom-probe tomography: {A}dvances and
  applications, Ann Rev Mater Res 37:127--158

\bibitem{Venables73PhilMag}
Venables JA, Harland CJ (1973) Electron back-scattering patterns---{A} new
  technique for obtaining crystallographic information in the scanning electron
  microscope, Phil Mag 27:1193--1200

\bibitem{Poulsen2004}
Poulsen HF (2004) Three-dimensional {X-ray} diffraction microscopy: {M}apping
  polycrystals and their dynamics, vol. 205, Springer Science \& Business Media

\bibitem{Horstemeyer2012}
Horstemeyer MF (2012) Integrated Computational Materials Engineering ({ICME})
  for Metals: {U}sing Multiscale Modeling to Invigorate Engineering Design with
  Science, Wiley-{TMS}, Hoboken, {N.J}

\bibitem{Eshelby57}
Eshelby JD (1957) The determination of the elastic field of an ellipsoidal
  inclusion, and related problems, P Roy Soc A-Math Phys 241:376--396

\bibitem{mura1987micromechanics}
Mura T (1987) Micromechanics of defects in solids, vol.~3, Springer Science \&
  Business Media

\bibitem{suquet1987elements}
Suquet P (1987) Elements of homogenization for inelastic solid mechanics, in:
  Sanchez-Palencia E, Zaoui A (eds.) Homogenization techniques for composite
  media, Springer-Verlag, pp. 193--278

\bibitem{ke2005experiments}
Ke CH, Pugno N, Peng B, Espinosa HD (2005) Experiments and modeling of carbon
  nanotube-based {NEMS} devices, J Mech Phys Solids 53:1314--1333

\bibitem{pelesko2002modeling}
Pelesko JA, Bernstein DH (2002) Modeling {MEMS} and {NEMS}, {CRC} Press

\bibitem{cao2009molecular}
Cao BY, Sun J, Chen M, Guo ZY (2009) Molecular momentum transport at
  fluid-solid interfaces in {MEMS/NEMS}: {A} review, Int J Mol Sci
  10:4638--4706

\bibitem{dorogin2013real}
Dorogin LM, Vlassov S, Polyakov B, Antsov M, L{\~o}hmus R, Kink I, Romanov AE
  (2013) Real-time manipulation of {ZnO} nanowires on a flat surface employed
  for tribological measurements: {E}xperimental methods and modeling, Phys
  Status Solidi B 250:305--317

\bibitem{zaumseil2003three}
Zaumseil J, Meitl MA, Hsu JWP, Acharya BR, Baldwin KW, Loo YL, Rogers JA (2003)
  Three-dimensional and multilayer nanostructures formed by nanotransfer
  printing, Nano Lett 3:1223--1227

\bibitem{pan2012hierarchical}
Pan L, Yu G, Zhai D, Lee HR, Zhao W, Liu N, Wang H, Tee BCK, Shi Y, Cui Y,
  et~al. (2012) Hierarchical nanostructured conducting polymer hydrogel with
  high electrochemical activity, P Natl Acad Sci USA 109:9287--9292

\bibitem{vaezi2013review}
Vaezi M, Seitz H, Yang S (2013) A review on {3D} micro-additive manufacturing
  technologies, Int J Adv Manuf Tech 67:1721--1754

\bibitem{legros2008situ}
Legros M, Gianola DS, Hemker KJ (2008) \emph{In situ} {TEM} observations of
  fast grain-boundary motion in stressed nanocrystalline aluminum films, Acta
  Mater 56:3380--3393

\bibitem{wang2011situ}
Wang CM, Xu W, Liu J, Zhang JG, Saraf LV, Arey BW, Choi D, Yang ZG, Xiao J,
  Thevuthasan S, et~al. (2011) \emph{In situ} transmission electron microscopy
  observation of microstructure and phase evolution in a {SnO$_2$} nanowire
  during lithium intercalation, Nano Lett 11:1874--1880

\bibitem{rollett2007three}
Rollett AD, Lee SB, Campman R, Rohrer GS (2007) Three-dimensional
  characterization of microstructure by electron back-scatter diffraction, Ann
  Rev Mater Res 37:627--658

\bibitem{groeber2008framework}
Groeber M, Ghosh S, Uchic MD, Dimiduk DM (2008) A framework for automated
  analysis and simulation of {3D} polycrystalline microstructures.: {P}art 1:
  {S}tatistical characterization, Acta Mater 56:1257--1273

\bibitem{groeber2014dream}
Groeber M A and~Jackson MA (2014) {DREAM 3D: A} digital representation
  environment for the analysis of microstructure in {3D}, Integr Mater Manuf
  Innov 3:1--17

\bibitem{brandt2002multiscale}
Brandt A (2002) Multiscale scientific computation: {R}eview 2001, in:
  {Multiscale and Multiresolution Methods}, Springer, pp. 3--95

\bibitem{E07}
Weinan E, Engquist B, Li X, Ren W (2007) Heterogeneous multiscale methods: {A}
  review, Commun Comput Phys 2:367--450

\bibitem{barenblatt1996scaling}
Barenblatt GI (1996) Scaling, self-similarity, and intermediate asymptotics:
  {D}imensional analysis and intermediate asymptotics, vol.~14, Cambridge
  University Press

\bibitem{cheng2004scaling}
Cheng YT, Cheng CM (2004) Scaling, dimensional analysis, and indentation
  measurements, Mater Sci Eng R 44:91--149

\bibitem{miguel2001intermittent}
Miguel MC, Vespignani A, Zapperi S, Weiss J, Grasso JR (2001) Intermittent
  dislocation flow in viscoplastic deformation, Nature 410:667--671

\bibitem{Csikor12102007}
Csikor FF, Motz C, Weygand D, Zaiser M, Zapperi S (2007) Dislocation
  avalanches, strain bursts, and the problem of plastic forming at the
  micrometer scale, Science 318:251--254

\bibitem{chen2010bending}
Chen YS, Choi W, Papanikolaou S, Sethna JP (2010) Bending crystals: {E}mergence
  of fractal dislocation structures, Phys Rev Lett 105:105501

\bibitem{bai2005statistical}
Bai YL, Wang HY, Xia MF, Ke FJ (2005) Statistical mesomechanics of solid,
  linking coupled multiple space and time scales, Appl Mech Rev 58:372--388

\bibitem{liu2013computational}
Liu Y, Greene MS, Chen W, Dikin DA, Liu WK (2013) Computational microstructure
  characterization and reconstruction for stochastic multiscale material
  design, Comput Aided Design 45:65--76

\bibitem{bachmann2011grain}
Bachmann F, Hielscher R, Schaeben H (2011) Grain detection from {2D} and {3D}
  {EBSD} data--{S}pecification of the {MTEX} algorithm, Ultramicroscopy
  111:1720--1733

\bibitem{germain2014identification}
Germain L, Kratsch D, Salib M, Gey N (2014) Identification of sub-grains and
  low angle boundaries beyond the angular resolution of {EBSD} maps, Mater
  Charact 98:66--72

\bibitem{bjorstad1986iterative}
Bj{\o}rstad PE, Widlund OB (1986) Iterative methods for the solution of
  elliptic problems on regions partitioned into substructures, {SIAM J Numer
  Anal} 23:1097--1120

\bibitem{pothen1990partitioning}
Pothen A, Simon HD, Liou KP (1990) Partitioning sparse matrices with
  eigenvectors of graphs, SIAM J Matrix Anal A 11:430--452

\bibitem{karypis1998fast}
Karypis G, Kumar V (1998) A fast and high quality multilevel scheme for
  partitioning irregular graphs, SIAM J Sci Comput 20:359--392

\bibitem{krysl2001dimensional}
Krysl P, Lall S, Marsden JE (2001) Dimensional model reduction in non-linear
  finite element dynamics of solids and structures, Int J Numer Meth Eng
  51:479--504

\bibitem{willcox2002balanced}
Willcox K, Peraire J (2002) Balanced model reduction via the proper orthogonal
  decomposition, AIAA Journal 40:2323--2330

\bibitem{bouvard2009review}
Bouvard JL, Ward DK, Hossain D, Nouranian S, Marin EB, Horstemeyer MF (2009)
  Review of hierarchical multiscale modeling to describe the mechanical
  behavior of amorphous polymers, J Eng Mater Tech 131:041206

\bibitem{li2013challenges}
Li Y, Abberton BC, Kr{\"o}ger M, Liu WK (2013) Challenges in multiscale
  modeling of polymer dynamics, Polymers 5:751--832

\bibitem{de2007multiscale}
de~Pablo JJ, Curtin WA (2007) Multiscale modeling in advanced materials
  research: {C}hallenges, novel methods, and emerging applications, MRS Bull
  32:905--911

\bibitem{praprotnik2008multiscale}
Praprotnik M, Site LD, Kremer K (2008) Multiscale simulation of soft matter:
  {F}rom scale bridging to adaptive resolution, Annu Rev Phys Chem 59:545--571

\bibitem{murtola2009multiscale}
Murtola T, Bunker A, Vattulainen I, Deserno M, Karttunen M (2009) Multiscale
  modeling of emergent materials: {B}iological and soft matter, Phys Chem Chem
  Phys 11:1869--1892

\bibitem{DeHoff1968}
DeHoff RT, Rhines FN (1968) Quantitative microscopy, {McGraw}-Hill

\bibitem{Fultz2012}
Fultz B, Howe JM (2012) Transmission Electron Microscopy and Diffractometry of
  Materials, Springer, Heidelberg

\bibitem{den2013estimation}
den Dekker A, Gonnissen J, De~Backer A, Sijbers J, Van~Aert S (2013) Estimation
  of unknown structure parameters from high-resolution {(S)TEM} images: {W}hat
  are the limits?, Ultramicroscopy 134:34--43

\bibitem{Watanabe1999}
Watanabe M, Williams DB (1999) Atomic-level detection by {X}-ray microanalysis
  in the analytical electron microscope, Ultramicroscopy 78:89--101

\bibitem{Genc2009}
Genc A, Banerjee R, Thompson GB, Maher DM, Johnson AW, Fraser HL (2009)
  Complementary techniques for the characterization of thin film {Ti/Nb}
  multilayers, Ultramicroscopy 109:1276--1281

\bibitem{Zhou2012}
Zhou W, Wachs IE, Kiely CJ (2012) Nanostructural and chemical characterization
  of supported metal oxide catalysts by aberration corrected analytical
  electron microscopy, Curr Opin Solid State Mater Sci 16:10--22

\bibitem{Blavette1993}
Blavette D, Bostel A, Sarrau J, Deconihout B, Menand A (1993) An atom probe for
  three-dimensional tomography, Nature 363:432--435

\bibitem{Marquis2013}
Marquis EA, Bachhav M, Chen Y, Dong Y, Gordon LM, Joester D, McFarland A (2013)
  On the current role of atom probe tomography in materials characterization
  and materials science, Curr Opin Solid State Mater Sci 17:217--223

\bibitem{Marquis2014}
Marquis EA, Bachhav M, Chen Y, Dong Y, Gordon LM, Joester D, McFarland A (2015)
  Corrigendum to ``{O}n the current role of atom probe tomography in materials
  characterization and materials science'' [{Current Opinion Solid State Mater.
  Sci.} 17/5 (2014) 217--223], Curr Opin Solid State Mater Sci 19:147

\bibitem{Sauvage2014}
Sauvage X, Enikeev N, Valiev R, Nasedkina Y, Murashkin M (2014) Atomic-scale
  analysis of the segregation and precipitation mechanisms in a severely
  deformed {A}l--{M}g alloy, Acta Mater 72:125--136

\bibitem{Bernier2011}
Bernier JV, Barton NR, Lienert U, Miller MP (2011) Far-field high-energy
  diffraction microscopy: {A} tool for intergranular orientation and strain
  analysis, J Strain Anal Eng 46:527--547

\bibitem{reischig2013advances}
Reischig P, King A, Nervo L, Vigano N, Guilhem Y, Palenstijn WJ, Batenburg KJ,
  Preuss M, Ludwig W (2013) Advances in {X}-ray diffraction contrast
  tomography: {F}lexibility in the setup geometry and application to multiphase
  materials, J Appl Crystallogr 46:297--311

\bibitem{Maire2014}
Maire E, Withers PJ (2014) Quantitative {X}-ray tomography, Int Mater Rev
  59:1--43

\bibitem{Schuren14}
Schuren JC, Shade PA, Bernier JV, Li SF, Blank B, Lind J, Kenesei P, Lienert U,
  Suter RM, Turner TJ, Dimiduk DM, Almer J (2014) New opportunities for
  quantitative tracking of polycrystal responses in three dimensions, Curr Opin
  Solid State Mater Sci 19:235--244

\bibitem{van2013deformation}
Van~Petegem S, Li L, Anderson PM, Van~Swygenhoven H (2013) Deformation
  mechanisms in nanocrystalline metals: {I}nsights from \emph{in-situ}
  diffraction and crystal plasticity modelling, Thin Solid Films 530:20--24

\bibitem{obstalecki2014quantitative}
Obstalecki M, Wong SL, Dawson PR, Miller MP (2014) Quantitative analysis of
  crystal scale deformation heterogeneity during cyclic plasticity using
  high-energy {X}-ray diffraction and finite-element simulation, Acta Mater
  75:259--272

\bibitem{lentz2015situ}
Lentz M, Klaus M, Beyerlein IJ, Zecevic M, Reimers W, Knezevic M (2015)
  \emph{In situ} {X}-ray diffraction and crystal plasticity modeling of the
  deformation behavior of extruded {Mg--Li--(Al)} alloys: {A}n uncommon
  tension--compression asymmetry, Acta Mater 86:254--268

\bibitem{clark2013ultrafast}
Clark JN, Beitra L, Xiong G, Higginbotham A, Fritz DM, Lemke HT, Zhu D, Chollet
  M, Williams GJ, Messerschmidt M, et~al. (2013) Ultrafast three-dimensional
  imaging of lattice dynamics in individual gold nanocrystals, Science
  341:56--59

\bibitem{milathianaki2013femtosecond}
Milathianaki D, Boutet S, Williams GJ, Higginbotham A, Ratner D, Gleason AE,
  Messerschmidt M, Seibert MM, Swift D, Hering P, et~al. (2013) Femtosecond
  visualization of lattice dynamics in shock-compressed matter, Science
  342:220--223

\bibitem{Golovin2008}
Golovin YI (2008) Nanoindentation and mechanical properties of solids in
  submicrovolumes, thin near-surface layers, and films: {A} review, Phys Solid
  State 50:2205--2236

\bibitem{lodes2011influence}
Lodes MA, Hartmaier A, G{\"o}ken M, Durst K (2011) Influence of dislocation
  density on the pop-in behavior and indentation size effect in {CaF}$_2$
  single crystals: {E}xperiments and molecular dynamics simulations, Acta Mater
  59:4264--4273

\bibitem{begau2011atomistic}
Begau C, Hartmaier A, George EP, Pharr GM (2011) Atomistic processes of
  dislocation generation and plastic deformation during nanoindentation, Acta
  Mater 59:934--942

\bibitem{ruestes2014atomistic}
Ruestes CJ, Stukowski A, Tang Y, Tramontina DR, Erhart P, Remington BA,
  Urbassek HM, Meyers MA, Bringa EM (2014) Atomistic simulation of tantalum
  nanoindentation: {E}ffects of indenter diameter, penetration velocity, and
  interatomic potentials on defect mechanisms and evolution, Mater Sci Eng A
  613:390--403

\bibitem{zambaldi2012orientation}
Zambaldi C, Yang Y, Bieler TR, Raabe D (2012) Orientation informed
  nanoindentation of $\alpha$-titanium: {I}ndentation pileup in hexagonal
  metals deforming by prismatic slip, J Mater Res 27:356--367

\bibitem{selvarajou2014orientation}
Selvarajou B, Shin JH, Ha TK, Choi IS, Joshi SP, Han HN (2014)
  Orientation-dependent indentation response of magnesium single crystals:
  {M}odeling and experiments, Acta Mater 81:358--376

\bibitem{Yao2014}
Yao WZ, Krill~{III} CE, Albinski B, Schneider HC, You JH (2014) Plastic
  material parameters and plastic anisotropy of tungsten single crystal: {A}
  spherical micro-indentation study, J Mater Sci 49:3705--3715

\bibitem{Minor2001}
Minor AM, Morris JWJ, Stach EA (2001) Quantitative \emph{in situ}
  nanoindentation in an electron microscope, Appl Phys Lett 79:1625--1627

\bibitem{legros2010quantitative}
Legros M, Gianola DS, Motz C (2010) Quantitative \emph{in situ} mechanical
  testing in electron microscopes, MRS Bull 35:354--360

\bibitem{kiener2011situ}
Kiener D, Hosemann P, Maloy SA, Minor AM (2011) \emph{In situ} nanocompression
  testing of irradiated copper, Nat Mater 10:608--613

\bibitem{ohmura2012effects}
Ohmura T, Zhang L, Sekido K, Tsuzaki K (2012) Effects of lattice defects on
  indentation-induced plasticity initiation behavior in metals, J Mater Res
  27:1742--1749

\bibitem{issa2015situ}
Issa I, Amodeo J, R{\'e}thor{\'e} J, Joly-Pottuz L, Esnouf C, Morthomas J,
  Perez M, Chevalier J, Masenelli-Varlot K (2015) In situ investigation of
  {MgO} nanocube deformation at room temperature, Acta Mater 86:295--304

\bibitem{liu2015situ}
Liu Y, Li N, Bufford D, Lee JH, Wang J, Wang H, Zhang X (2015) In situ
  nanoindentation studies on detwinning and work hardening in nanotwinned
  monolithic metals, {JOM} :1--9

\bibitem{soer2004effects}
Soer WA, De~Hosson JTM, Minor AM, Morris JW, Stach EA (2004) Effects of solute
  mg on grain boundary and dislocation dynamics during nanoindentation of
  al--mg thin films, Acta Mater 52:5783--5790

\bibitem{Gouldstone2007}
Gouldstone A, Chollacoop N, Dao M, Li J, Minor AM, Shen YL (2007) Indentation
  across size scales and disciplines: {R}ecent developments in experimentation
  and modeling, Acta Mater 55:4015--4039

\bibitem{bufford2014situ}
Bufford D, Liu Y, Wang J, Wang H, Zhang X (2014) In situ nanoindentation study
  on plasticity and work hardening in aluminium with incoherent twin
  boundaries, Nat Commun 5

\bibitem{Boehler1987}
Boehler J, Elaoufi L, Raclin J (1987) On experimental testing methods for
  anisotropic materials, Res Mech 21:73--95

\bibitem{Rey1984}
Bretheau T, Mussot P, Rey C (1984) Microscale plastic inhomogeneities and
  macroscopic behavior of single and multiphase materials, J Eng Mater-T ASME
  106:304--310

\bibitem{Allais1994}
Allais L, Bornert M, Bretheau T, Caldemaison D (1994) Experimental
  characterization of the local strain field in a heterogeneous elastoplastic
  material, Acta Metall Mater 42:3865--3880

\bibitem{Rey1997}
Rey C, DeLesegno PV (1997) Experimental analysis of bifurcation and post
  bifurcation in iron single crystals, Mater Sci Eng A 234:1007--1010

\bibitem{Martin2013}
Martin G, Sinclair CW, Schmitt JH (2013) Plastic strain heterogeneities in an
  {Mg--1Zn--0.5Nd} alloy, Scripta Mater 68:695--698

\bibitem{Raphanel2000}
Delaire F, Raphanel JL, Rey C (2000) Plastic heterogeneities of a copper
  multicrystal deformed in uniaxial tension: {E}xperimental study and finite
  element simulations, Acta Mater 48:1075--1087

\bibitem{soppa2001experimental}
Soppa E, Doumalin P, Binkele P, Wiesendanger T, Bornert M, Schmauder S (2001)
  Experimental and numerical characterisation of in-plane deformation in
  two-phase materials, Comp Mater Sci 21:261--275

\bibitem{Heripre07}
Heripre E, Dexet M, Crepin J, G\'elebart L, Roos A, Bornert M, Caldemaison D
  (2007) Coupling between experimental measurements and polycrystal finite
  element calculations for micromechanical study of metallic materials, Int J
  Plasticity 23:1512--1539

\bibitem{Bornert2000}
Doumalin P, Bornert M (2000) Micromechanical applications of digital image
  correlation techniques, in: {Interferometry in Speckle Light: Theory and
  Applications}, pp. 67--74

\bibitem{Pippan2011}
Rehrl C, Kleber S, Antretter T, Pippan R (2011) A methodology to study crystal
  plasticity inside a compression test sample based on image correlation and
  {EBSD}, Mater Charact 62:793--800

\bibitem{Celotto2012}
Ghadbeigi H, Pinna C, Celotto S (2012) Quantitative strain analysis of the
  large deformation at the scale of microstructure: {C}omparison between
  digital image correlation and microgrid techniques, Exp Mech 52:1483--1492

\bibitem{vanderesse2013open}
Vanderesse N, Lagac{\'e} M, Bridier F, Bocher P (2013) An open source software
  for the measurement of deformation fields by means of digital image
  correlation, Microsc Microanal 19:820--821

\bibitem{Tasan2014}
Tasan CC, Diehl M, Yan D, Zambaldi C, Shanthraj P, Roters F, Raabe D (2014)
  Integrated experimental--simulation analysis of stress and strain
  partitioning in multiphase alloys, Acta Mater 81:386--400

\bibitem{stinville2015high}
Stinville JC, Vanderesse N, Bridier F, Bocher P, Pollock TM (2015) High
  resolution mapping of strain localization near twin boundaries in a
  nickel-based superalloy, Acta Mater 98:29--42

\bibitem{Adams1993}
Adams B, Wright S, Kunze K (1993) Orientation imaging: {T}he emergence of a new
  microscopy, Metall Trans A 24:819--831

\bibitem{Zaefferer2003}
Thomas I, Zaefferer S, Friedel F, Raabe D (2003) High-resolution {EBSD}
  investigation of deformed and partially recrystallized {IF} steel, Adv Eng
  Mater 5:566--570

\bibitem{Rollett2009}
Mishra SK, Pant P, Narasimhan K, Rollett AD, Samajdar I (2009) On the widths of
  orientation gradient zones adjacent to grain boundaries, Scripta Mater
  61:273--276

\bibitem{Vignal2011}
Clair A, Foucault M, Calonne O, Lacroute Y, Markey L, Salazar M, Vignal V,
  Finot E (2011) Strain mapping near a triple junction in strained {Ni}-based
  alloy using {EBSD} and biaxial nanogauges, Acta Mater 59:3116--3123

\bibitem{Bacroix2013}
Jedrychowski M, Tarasiuk J, Bacroix B, Wronski S (2013) Electron backscatter
  diffraction investigation of local misorientations and orientation gradients
  in connection with evolution of grain boundary structures in deformed and
  annealed zirconium. {A} new approach in grain boundary analysis, J Appl
  Crystallogr 46:483--492

\bibitem{Jimenez2015}
Cepeda-Jimenez CM, Molina-Aldareguia JM, Perez-Prado MT (2015) Effect of grain
  size on slip activity in pure magnesium polycrystals, Acta Mater 84:443--456

\bibitem{Zaefferer2008}
Zaefferer S, Romano P, Friedel F (2008) {EBSD} as a tool to identify and
  quantify bainite and ferrite in low-alloyed {Al-TRIP} steels, J
  Microsc-Oxford 230:499--508

\bibitem{calcagnotto2010orientation}
Calcagnotto M, Ponge D, Demir E, Raabe D (2010) Orientation gradients and
  geometrically necessary dislocations in ultrafine grained dual-phase steels
  studied by {2D and 3D EBSD}, Mater Sci Eng A 527:2738--2746

\bibitem{Wilkinson2010}
Wilkinson AJ, Clarke EE, Britton TB, Littlewood P, Karamched PS (2010)
  High-resolution electron backscatter diffraction: {A}n emerging tool for
  studying local deformation, J Strain Anal Eng 45:365--376

\bibitem{ruggles2013estimations}
Ruggles TJ, Fullwood DT (2013) Estimations of bulk geometrically necessary
  dislocation density using high resolution {EBSD}, Ultramicroscopy 133:8--15

\bibitem{liang2009gnd}
Liang H, Dunne FPE (2009) {GND} accumulation in bi-crystal deformation:
  {C}rystal plasticity analysis and comparison with experiments, Int J Mech Sci
  51:326--333

\bibitem{Fressengeas2013}
Beausir B, Fressengeas C (2013) Disclination densities from {EBSD} orientation
  mapping, Int J Solids Struct 50:137--146

\bibitem{bingham2010statistical}
Bingham MA, Lograsso BK, Laabs FC (2010) A statistical analysis of the
  variation in measured crystal orientations obtained through electron
  backscatter diffraction, Ultramicroscopy 110:1312--1319

\bibitem{beyerlein2010statistical}
Beyerlein IJ, Capolungo L, Marshall PE, McCabe RJ, Tom{\'e} CN (2010)
  Statistical analyses of deformation twinning in magnesium, Phil Mag
  90:2161--2190

\bibitem{juan2015statistical}
Juan PA, Pradalier C, Berbenni S, McCabe RJ, Tom{\'e} CN, Capolungo L (2015) A
  statistical analysis of the influence of microstructure and twin--twin
  junctions on twin nucleation and twin growth in {Z}r, Acta Mater 95:399--410

\bibitem{Bacroix2005}
Gerber PH, Tarasiuk J, Chiron R, Bacroix B (2005) Estimation of the
  recrystallized volume fraction from local misorientation calculations, Arch
  Metall Mater 50:747--755

\bibitem{Zaefferer2000}
Zaefferer S (2000) New developments of computer-aided crystallographic analysis
  in transmission electron microscopy, J Appl Crystallogr 33:10--25

\bibitem{Rauch2014}
Rauch EF, Veron M (2014) Automated crystal orientation and phase mapping in
  {TEM}, Mater Charact 98:1--9

\bibitem{Taheri2015}
Leff AC, Weinberger CR, Taheri ML (2015) Estimation of dislocation density from
  precession electron diffraction data using the {N}ye tensor, Ultramicroscopy
  153:9--21

\bibitem{Hytch1998}
Hytch MJ, Snoeck E, Kilaas R (1998) Quantitative measurement of displacement
  and strain fields from {HREM} micrographs, Ultramicroscopy 74:131--146

\bibitem{Hytch2014}
Lubk A, Javon E, Cherkashin N, Reboh S, Gatel C, Hytch M (2014) Dynamic
  scattering theory for dark-field electron holography of 3{D} strain fields,
  Ultramicroscopy 136:42--49

\bibitem{Beche2013}
B\'ech\'e A, Rouvi\`ere JL, Barnes JP, Cooper D (2013) Strain measurement at
  the nanoscale: {C}omparison between convergent beam electron diffraction,
  nano-beam electron diffraction, high resolution imaging and dark field
  electron holography, Ultramicroscopy 131:10--23

\bibitem{Rowenhorst2006}
Rowenhorst DJ, Gupta A, Feng CR, Spanos G (2006) {3D} crystallographic and
  morphological analysis of coarse martensite: {C}ombining {EBSD} and serial
  sectioning, Scripta Mater 55:11--16

\bibitem{Thebault2008}
Thebault J, Solas D, Rey C, Baudin T, Fandeur O, Clavel M (2008)
  Polycrystalline modelling of {Udimet} 720 forging, in: {Superalloys 2008},
  {TMS}, pp. 985--992

\bibitem{Cedat2012}
C\'edat D, Fandeur O, Rey C, Raabe D (2012) Polycrystal model of the mechanical
  behavior of a {Mo}--{TiC} 30 vol.\% metal--ceramic composite using a
  three-dimensional microstructure map obtained by dual beam focused ion beam
  scanning electron microscopy, Acta Mater 60:1623--1632

\bibitem{Uchic2006}
Uchic MD, Groeber MA, Dimiduk DM, Simmons JP (2006) {3D} microstructural
  characterization of nickel superalloys via serial-sectioning using a dual
  beam {FIB-SEM}, Scripta Mater 55:23--28

\bibitem{endo2014three}
Endo T, Sugino Y, Ohono N, Ukai S, Miyazaki N, Wang Y, Ohnuki S (2014)
  Three-dimensional characterization of {ODS} ferritic steel using by {FIB-SEM}
  serial sectioning method, Microscopy 63:i23--i23

\bibitem{yamasaki20153d}
Yamasaki S, Mitsuhara M, Ikeda K, Hata S, Nakashima H (2015) {3D} visualization
  of dislocation arrangement using scanning electron microscope serial
  sectioning method, Scripta Mater 101:80--83

\bibitem{Bernacki2010}
Log{\'e} R, Resk H, Sun Z, Delannay L, Bernacki M (2010) Modeling of plastic
  deformation and recrystallization of polycrystals using digital
  microstructures and adaptive meshing techniques, Steel Res Int 81:1420--1425

\bibitem{Buffiere2006}
Baruchel J, Buffiere JY, Cloetens P, Di~Michiel M, Ferrie E, Ludwig W, Maire E,
  Salvo L (2006) Advances in synchrotron radiation microtomography, Scripta
  Mater 55:41--46

\bibitem{Evrard2010}
Evrard P, El~Bartali A, Aubin V, Rey C, Degallaix S, Kondo D (2010) Influence
  of boundary conditions on bi-phased polycrystal microstructure calculation,
  Int J Solids Struct 47:1979--1986

\bibitem{oddershede2012measuring}
Oddershede J, Camin B, Schmidt S, Mikkelsen LP, S{\o}rensen HO, Lienert U,
  Poulsen HF, Reimers W (2012) Measuring the stress field around an evolving
  crack in tensile deformed {Mg AZ}31 using three-dimensional {X}-ray
  diffraction, Acta Mater 60:3570--3580

\bibitem{Withers12}
Withers PJ, Preuss M (2012) Fatigue and damage in structural materials studied
  by {X}-ray tomography, Ann Rev Mater Res 42:81--103

\bibitem{poulsen2012introduction}
Poulsen HF (2012) An introduction to three-dimensional {X}-ray diffraction
  microscopy, J Appl Crystallogr 45:1084--1097

\bibitem{chow2014measurement}
Chow W, Solas D, Puel G, Perrin E, Baudin T, Aubin V (2014) Measurement of
  complementary strain fields at the grain scale, in: Adv Mater Res, Trans Tech
  Publ, vol. 996, pp. 64--69

\bibitem{miller2014understanding}
Miller MP, Dawson PR (2014) Understanding local deformation in metallic
  polycrystals using high energy {X}-rays and finite elements, Curr Opin Solid
  State Mater Sci 18:286--299

\bibitem{Pokharel14}
Pokharel R, Lind J, Kanjarla AK, Lebensohn RA, Li SF, Kenesei P, Suter RM,
  Rollett AD (2014) Polycrystal plasticity: {C}omparison between grain --
  {S}cale observations of deformation and simulations, Ann Rev Cond Mat Phys
  5:317--346

\bibitem{olsson2015strain}
Olsson CO, Bostr{\"o}m M, Buslaps T, Steuwer A (2015) Strain profiling of a
  ferritic-martensitic stainless steel sheet--{C}omparing synchrotron with
  conventional {X}-{R}ay {D}iffraction, Strain 51:71--77

\bibitem{Rollett2015}
Pokharel R, Lind J, Li SF, Kenesei P, Lebensohn RA, Suter RM, Rollett AD (2015)
  \emph{In-situ} observation of bulk {3D} grain evolution during plastic
  deformation in polycrystalline {Cu}, Int J Plasticity 67:217--234

\bibitem{Spear14}
Spear AD, Li SF, Lind JF, Suter RM, Ingraffea AR (2014) Three-dimensional
  characterization of microstructurally small fatigue-crack evolution using
  quantitative fractography combined with post-mortem {X-ray} tomography and
  high-energy {X}-ray diffraction microscopy, Acta Mater 76:413--424

\bibitem{Bienvenu2012}
Burteau A, N'Guyen F, Bartout JD, Forest S, Bienvenu Y, Saberi S, Naumann D
  (2012) Impact of material processing and deformation on cell morphology and
  mechanical behavior of polyurethane and nickel foams, Int J Solids Struct
  49:2714--2732

\bibitem{jornsanoh2011electron}
Jornsanoh P, Thollet G, Ferreira J, Masenelli-Varlot K, Gauthier C, Bogner A
  (2011) Electron tomography combining {ESEM} and {STEM}: {A} new {3D} imaging
  technique, Ultramicroscopy 111:1247--1254

\bibitem{masenelli2014wet}
Masenelli-Varlot K, Malch{\`e}re A, Ferreira J, Heidari~Mezerji Hand~Bals S,
  Messaoudi C, Marco~Garrido S (2014) Wet-{STEM} tomography: {P}rinciples,
  potentialities and limitations, Microsc and Microanal 20:366--375

\bibitem{browning2012recent}
Browning ND, Bonds MA, Campbell GH, Evans JE, LaGrange T, Jungjohann KL, Masiel
  DJ, McKeown J, Mehraeen S, Reed BW, et~al. (2012) Recent developments in
  dynamic transmission electron microscopy, Curr Opin Solid State Mater Sci
  16:23--30

\bibitem{flannigan20124d}
Flannigan DJ, Zewail AH (2012) {4D} electron microscopy: {P}rinciples and
  applications, Acc Chem Res 45:1828--1839

\bibitem{baum2014towards}
Baum P (2014) Towards ultimate temporal and spatial resolutions with ultrafast
  single-electron diffraction, J Phys B 47:124005

\bibitem{kelly2004first}
Kelly TF, Gribb T, Olson JD, Martens RL, Shepard JD, Wiener SA, Kunicki TC,
  Ulfig RM, Lenz DR, Strennen EM, et~al. (2004) First data from a commercial
  local electrode atom probe ({LEAP}), Microsc Microanal 10:373--383

\bibitem{isheim2006atom}
Isheim D, Kolli RP, Fine ME, Seidman DN (2006) An atom-probe tomographic study
  of the temporal evolution of the nanostructure of {F}e--{C}u based
  high-strength low-carbon steels, Scripta Mater 55:35--40

\bibitem{kelly2007atom}
Kelly TF, Miller MK (2007) Atom probe tomography, Rev Sci Inst 78:031101

\bibitem{miller2012future}
Miller MK, Kelly TF, Rajan K, Ringer SP (2012) The future of atom probe
  tomography, Mater Today 15:158--165

\bibitem{lagrange2012approaches}
LaGrange T, Reed BW, Santala MK, McKeown JT, Kulovits A, Wiezorek JMK, Nikolova
  L, Rosei F, Siwick BJ, Campbell GH (2012) Approaches for ultrafast imaging of
  transient materials processes in the transmission electron microscope, Micron
  43:1108--1120

\bibitem{kalidindi2015materials}
Kalidindi SR, De~Graef M (2015) Materials data science: {C}urrent status and
  future outlook, Annu Rev Mater Res 45:171--93

\bibitem{Molinari97}
Molinari A, Ahzi S, Kouddane R (1997) On the self-consistent modeling of
  elastic-plastic behavior of polycrystals, Mech Mater 26:43--62

\bibitem{Lebensohn97}
Lebensohn RA, Canova GR (1997) A self-consistent approach for modelling texture
  development of two-phase polycrystals: {A}pplication to titanium alloys, Acta
  Mater 45:3687--3694

\bibitem{Berbenni07}
Berbenni S, Favier V, Berveiller M (2007) Impact of the grain size distribution
  on the yield stress of heterogeneous materials, Int J Plasticity 23:114--142

\bibitem{Zeghadi07-2}
Zeghadi A, Forest S, Gourgues AF, Bouaziz O (2007) Ensemble averaging
  stress--strain fields in polycrystalline aggregates with a constrained
  surface microstructure--{P}art 2: {C}rystal plasticity, Phil Mag
  87:1425--1446

\bibitem{Schroder14}
Schroder J (2014) A numerical two-scale homogenization scheme: {T}he
  {FE}$^2$-method, in: Schroder J, Hackl K (eds.) Plasticity and Beyond,
  Springer Vienna, vol. 550 of \emph{CISM International Centre for Mechanical
  Sciences}, pp. 1--64

\bibitem{Cailletaud03}
Cailletaud G, Forest S, Jeulin D, Feyel F, Galliet I, Mounoury V, Quilici S
  (2003) Some elements of microstructural mechanics, Comp Mater Sci 27:351--374

\bibitem{Li10}
Li Z, Wen B, Zabaras N (2010) Computing mechanical response variability of
  polycrystalline microstructures through dimensionality reduction techniques,
  Comp Mater Sci 49:568--581

\bibitem{Guilleminot11}
Guilleminot J, Noshadravan A, Soize C, Ghanem RG (2011) A probabilistic model
  for bounded elasticity tensor random fields with application to
  polycrystalline microstructures, Comp Method Appl M 200:1637--1648

\bibitem{fullwood2008microstructure}
Fullwood DT, Niezgoda SR, Kalidindi SR (2008) Microstructure reconstructions
  from 2-point statistics using phase-recovery algorithms, Acta Mater
  56:942--948

\bibitem{xu2014descriptor}
Xu H, Dikin DA, Burkhart C, Chen W (2014) Descriptor-based methodology for
  statistical characterization and {3D} reconstruction of microstructural
  materials, Comp Mater Sci 85:206--216

\bibitem{turner2016statistical}
Turner DM, Kalidindi SR (2016) Statistical construction of 3-{D}
  microstructures from 2-{D} exemplars collected on oblique sections, Acta
  Mater 102:136--148

\bibitem{Altendorf14}
Altendorf H, Latourte F, Jeulin D, Faessel M, Saintoyant L (2014) {3D}
  reconstruction of a multiscale microstructure by anisotropic tessellation
  models, Image Anal Stereol 33:121--130

\bibitem{Li13}
Li SF, Suter RM (2013) Adaptive reconstruction method for three-dimensional
  orientation imaging, J Appl Crystallogr 46:512--524

\bibitem{Lieberman15}
Lieberman EJ, Rollett AD, Lebensohn RA, Kober EM (2015) Calculation of grain
  boundary normals directly from {3D} microstructure images, Model Simul Mater
  Sc 23:035005

\bibitem{HerreraSolaz14}
Herrera-Solaz V, LLorca J, Dogan E, Karaman I, Segurado J (2014) An inverse
  optimization strategy to determine single crystal mechanical behavior from
  polycrystal tests: {A}pplication to {AZ31} {Mg} alloy, Int J Plasticity
  57:1--15

\bibitem{Sab92}
Sab K (1992) On the homogenization and the simulation of random materials, Eur
  J Mech A-Solid 5:585--607

\bibitem{Bourgeat94}
Bourgeat A, Mikelic A, Wright S (1994) Stochastic two-scale convergence in the
  mean and applications, J Reine Angew Math 456:19--52

\bibitem{Cafferelli05}
Caffarelli LA, Souganidis PE, Wang L (2005) Homogenization of fully nonlinear,
  uniformly elliptic and parabolic partial differential equations in stationary
  ergodic media, Commun Pur Appl Math 58:319--361

\bibitem{Bourgeat04}
Bourgeat A, Piatnitski A (2004) Approximations of effective coefficients in
  stochastic homogenization, Ann I H Poincar{\'e}-Pr 40:153--165

\bibitem{Cottereau13}
Cottereau R (2013) Numerical strategy for unbiased homogenization of random
  materials, Int J Numer Meth Eng 95:71--90

\bibitem{Koutsourelakis07}
Koutsourelakis PS (2007) Stochastic upscaling in solid mechanics: {A}n exercise
  in machine learning, J Comput Phys 226:301--325

\bibitem{Soize08}
Soize C (2008) Tensor-valued random fields for meso-scale stochastic model of
  anisotropic elastic microstructure and probabilistic analysis of
  representative volume element size, Probabilist Eng Mech 23:307--323

\bibitem{Clement13}
Clement A, Soize C, Yvonnet J (2013) Uncertainty quantification in
  computational stochastic multiscale analysis of nonlinear elastic materials,
  Comp Method Appl M 254:61--82

\bibitem{Xu07}
Xu XF (2007) A multiscale stochastic finite element method on elliptic problems
  involving uncertainties, Comp Method Appl M 196:2723--2736

\bibitem{Chevreuil13}
Chevreuil M, Nouy A, Safatly E (2013) A multiscale method with patch for the
  solution of stochastic partial differential equations with localized
  uncertainties, Comp Method Appl M 255:255--274

\bibitem{BenDhia98}
Ben~Dhia H (1998) Multiscale mechanical problems: {T}he {A}rlequin method, Cr
  Acad Sci II B 326:899--904

\bibitem{Cottereau11}
Cottereau R, Clouteau D, Ben~Dhia H, Zaccardi C (2011) A
  stochastic-deterministic coupling method for continuum mechanics, Comp Method
  Appl M 200:3280--3288

\bibitem{LeGuennec14}
Le~Guennec Y, Cottereau R, Clouteau D, Soize C (2014) A coupling method for
  stochastic continuum models at different scales, Probabilist Eng Mech
  37:138--147

\bibitem{Puel14}
Puel G, Aubry D (2014) Efficient fatigue simulation using periodic
  homogenization with multiple time scales, Int J Multiscale Com 12:291--318

\bibitem{Fish12}
Fish J, Bailakanavar M, Powers L, Cook T (2012) Multiscale fatigue life
  prediction model for heterogeneous materials, Int J Numer Meth Eng
  91:1087--1104

\bibitem{Ren07}
Ren W (2007) Seamless multiscale modeling of complex fluids using fiber bundle
  dynamics, Commun Math Sci 5:1027--1037

\bibitem{E07-2}
Weinan E, Liu D, Vanden-Eijnden E (2007) Nested stochastic simulation
  algorithms for chemical kinetic systems with multiple time scales, J Comput
  Phys 221:158--180

\bibitem{Gupta2013}
Lao J, Tam MN, Pinisetty D, Gupta N (2013) Molecular dynamics simulation of
  {FCC} metallic nanowires: {A} review, {JOM} 65:175--184

\bibitem{Romig2003}
Romig~Jr AD, Dugger MT, McWhorter PJ (2003) Materials issues in
  microelectromechanical devices: {S}cience, engineering, manufacturability and
  reliability, Acta Mater 51:5837--5866

\bibitem{Lee2006}
Lee Z, Ophus C, Fischer LM, Nelson-Fitzpatrick N, Westra KL, Evoy S, Radmilovic
  V, Dahmen U, Mitlin D (2006) Metallic {NEMS} components fabricated from
  nanocomposite {Al--Mo} films, Nanotechnology 17:3063--3070

\bibitem{Shaw2000}
Shaw TM, Trolier-McKinstry S, McIntyre PC (2000) The properties of
  ferroelectric films at small dimensions, Ann Rev Mater Sci 30:263--298

\bibitem{Eberl2009}
Gianola DS, Eberl C (2009) Micro- and nanoscale tensile testing of materials,
  {JOM} 61:24--35

\bibitem{Ashby2010}
Fleck NA, Deshpande VS, Ashby MF (2010) Micro-architectured materials: {P}ast,
  present and future, P Roy Soc A-Math Phys 466:2495--2516

\bibitem{Ashby2013}
Ashby M (2013) Designing architectured materials, Scripta Mater 68:4--7

\bibitem{Embury2010}
Embury D, Bouaziz O (2010) Steel-based composites: {D}riving forces and
  classifications, Ann Rev Mater Res 40:213--241

\bibitem{Ramirez2011}
Ramirez DA, Murr LE, Li SJ, Tian YX, Martinez E, Martinez JL, Machado BI,
  Gaytan S Mand~Medina F, Wicker RB (2011) Open-cellular copper structures
  fabricated by additive manufacturing using electron beam melting, Mater Sci
  Eng A 528:5379--5386

\bibitem{Simonelli2014}
Simonelli M, Tse YY, Tuck C (2014) Effect of the build orientation on the
  mechanical properties and fracture modes of {SLM} {Ti--6Al--4V}, Mater Sci
  Eng A 616:1--11

\bibitem{Kumar2003}
Kumar KS, Van~Swygenhoven H, Suresh S (2003) Mechanical behavior of
  nanocrystalline metals and alloys, Acta Mater 51:5743--5774

\bibitem{Shimokawa2014}
Shimokawa T, Oguro T, Tanaka M, Higashida K, Ohashi T (2014) A multiscale
  approach for the deformation mechanism in pearlite microstructure:
  {A}tomistic study of the role of the heterointerface on ductility, Mater Sci
  Eng A 598:68--76

\bibitem{VanVliet2003}
Van~Vliet KJ, Li J, Zhu T, Yip S, Suresh S (2003) Quantifying the early stages
  of plasticity through nanoscale experiments and simulations, Phys Rev B 67

\bibitem{Clegg2014}
Di~Gioacchino F, Clegg WJ (2014) Mapping deformation in small-scale testing,
  Acta Mater 78:103--113

\bibitem{Pollock2013}
Pollock TM, LeSar R (2013) The feedback loop between theory, simulation and
  experiment for plasticity and property modeling, Curr Opin Solid State Mater
  Sci 17:10--18

\bibitem{phillips2011diffraction}
Phillips PJ, Brandes MC, Mills MJ, De~Graef M (2011) Diffraction contrast
  {STEM} of dislocations: {I}maging and simulations, Ultramicroscopy
  111:1483--1487

\bibitem{coleman2015atomistic}
Coleman SP, Spearot DE (2015) Atomistic simulation and virtual diffraction
  characterization of homophase and heterophase alumina interfaces, Acta Mater
  82:403--413

\bibitem{stukowski2009atomistic}
Stukowski A, Markmann J, Weissm{\"u}ller J, Albe K (2009) Atomistic origin of
  microstrain broadening in diffraction data of nanocrystalline solids, Acta
  Mater 57:1648--1654

\bibitem{bristowe1980atomic}
Bristowe PD, Sass SL (1980) The atomic structure of a large angle [001] twist
  boundary in gold determined by a joint computer modelling and {X}-ray
  diffraction study, Acta Metall 28:575--588

\bibitem{bristowe1984effect}
Bristowe PD, Balluffi RW (1984) Effect of secondary relaxations on diffraction
  from high-$\sigma$ [001] twist boundaries, Surf Sci 144:14--27

\bibitem{oh1986structural}
Oh Y, Vitek V (1986) Structural multiplicity of $\sigma$ = 5 (001) twist
  boundaries and interpretation of {X}-ray diffraction from these boundaries,
  Acta Metall 34:1941--1953

\bibitem{higginbotham2013molecular}
Higginbotham A, Suggit MJ, Bringa EM, Erhart P, Hawreliak JA, Mogni G, Park N,
  Remington BA, Wark JS (2013) Molecular dynamics simulations of shock-induced
  deformation twinning of a body-centered-cubic metal, Phys Rev B 88:104105

\bibitem{mogni2014molecular}
Mogni G, Higginbotham A, Ga{\'a}l-Nagy K, Park N, Wark JS (2014) Molecular
  dynamics simulations of shock-compressed single-crystal silicon, Phys Rev B
  89:064104

\bibitem{van2005situ}
Van~Swygenhoven H, Budrovi{\'c} {\v{Z}}, Derlet PM, Fr{\o}seth AG, Van~Petegem
  S (2005) In situ diffraction profile analysis during tensile deformation
  motivated by molecular dynamics, Mater Sci Eng A 400:329--333

\bibitem{leyssale2009image}
Leyssale JM, Da~Costa JP, Germain C, Weisbecker P, Vignoles GL (2009) An
  image-guided atomistic reconstruction of pyrolytic carbons, Appl Phys Lett
  95:231912

\bibitem{robertson2013structural}
Robertson AW, Montanari B, He K, Allen CS, Wu YA, Harrison NM, Kirkland AI,
  Warner JH (2013) Structural reconstruction of the graphene monovacancy, ACS
  nano 7:4495--4502

\bibitem{Picard14}
Picard YN, Liu M, Lammatao J, Kamaladasa R, De~Graef M (2014) Theory of
  dynamical electron channeling contrast images of near-surface crystal
  defects, Ultramicroscopy 146:71--78

\bibitem{Schlesinger1979}
Schlesinger S, Crosbie RE, Gagne RE, Innis GS, Lalwani CS, Loch J, Sylvester
  RJ, Wright RD, Kheir N, Bartos D (1979) Terminology for model credibility,
  Simulation 32:103--104

\bibitem{CICME2008}
{National Research Council} (2008) Integrated Computational Materials
  Engineering: {A} Transformational Discipline for Improved Competitiveness and
  National Security, The National Academies Press, Washington, {DC}

\bibitem{McDowell-2008}
McDowell D, Olson G (2008) Concurrent design of hierarchical materials and
  structures, Sci Model Simul 15:207--240

\bibitem{Thacker-2004}
Thacker BH, Doebling SW, Hemez FM, Anderson MC, Pepin JE, Rodriguez EA (2004)
  Concepts of Model Verification and Validation, Tech. Rep. LA-14167, Los
  Alamos National Lab

\bibitem{Oberkampf-2004}
Oberkampf WL, Trucano TG, Hirsch C (2004) Verification, validation, and
  predictive capability in computational engineering and physics, Appl Mech Rev
  57:345--384

\bibitem{Cowles-2012}
Cowles B, Backman D, Dutton R (2012) Verification and validation of {ICME}
  methods and models for aerospace applications, Integr Mater Manuf Innov
  1:1--16

\bibitem{Wei-2012}
Wei X, Kysar JW (2012) Experimental validation of multiscale modeling of
  indentation of suspended circular graphene membranes, Int J Solids Struct
  49:3201--3209

\bibitem{Pen-2011}
Pen HM, Liang YC, Luo XC, Bai QS, Goel S, Ritchie JM (2011) Multiscale
  simulation of nanometric cutting of single crystal copper and its
  experimental validation, Comp Mater Sci 50:3431--3441

\bibitem{Gates-2005}
Gates TS, Odegard GM, Frankland SJV, Clancy TC (2005) Computational materials:
  {M}ulti-scale modeling and simulation of nanostructured materials, Compos Sci
  Techno 65:2416--2434

\bibitem{Fan2011}
Fan J (2011) Multiscale Analysis of Deformation and Failure of Materials, Wiley

\bibitem{plimpton1995fast}
Plimpton S (1995) Fast parallel algorithms for short-range molecular dynamics,
  J Comput Phys 117:1--19

\bibitem{McDowell2003}
McDowell DL, Gall K, Horstemeyer MF, Fan J (2003) Microstructure-based fatigue
  modeling of cast {A356-T6} alloy, Eng Fract Mech 70:49--80

\bibitem{Horstemeyer2010}
Horstemeyer MF (2010) Multiscale modeling: {A} review, in: Leszczynski J,
  Shukla MK (eds.) Practical Aspects of Computational Chemistry, Springer, pp.
  87--135

\bibitem{Chawla2005}
Chawla N, Deng X (2005) Microstructure and mechanical behavior of porous
  sintered steels, Mater Sci Eng A 390:98--112

\bibitem{Roberts2001}
Roberts AP, Garboczi EJ (2001) Elastic moduli of model random three-dimensional
  closed-cell cellular solids, Acta Mater 49:189--197

\bibitem{Ghosh2004}
Ghosh S, Moorthy S (2004) Three dimensional voronoi cell finite element model
  for microstructures with ellipsoidal heterogeneties, Comput Mech 34:510--531

\bibitem{Chawla2004}
Chawla N, Ganesh VV, Wunsch B (2004) Three-dimensional ({3D}) microstructure
  visualization and finite element modeling of the mechanical behavior of {SiC}
  particle reinforced aluminum composites, Scripta Mater 51:161--165

\bibitem{Lewis2006}
Lewis AC, Geltmacher AB (2006) Image-based modeling of the response of
  experimental {3D} microstructures to mechanical loading, Scripta Mater
  55:81--85

\bibitem{Maire2003}
Maire E, Fazekas A, Salvo L, Dendievel R, Youssef S, Cloetens P, Letang JM
  (2003) {X}-ray tomography applied to the characterization of cellular
  materials. {R}elated finite element modeling problems, Compos Sci Techno
  63:2431--2443

\bibitem{Youssef2005}
Youssef S, Maire E, Gaertner R (2005) Finite element modelling of the actual
  structure of cellular materials determined by {X}-ray tomography, Acta Mater
  53:719--730

\bibitem{Kenesei2004}
Kenesei P, Borb\'{e}ly A, Biermann H (2004) Microstructure based
  three-dimensional finite element modeling of particulate reinforced
  metal--matrix composites, Mater Sci Eng {A} 387--389:852--856

\bibitem{Larson2013236}
Larson DJ, Gault B, Geiser BP, De~Geuser F, Vurpillot F (2013) Atom probe
  tomography spatial reconstruction: {S}tatus and directions, Curr Opin Solid
  State Mater Sci 17:236--247

\bibitem{Prakash-2015}
Prakash A, Gu{\'e}nol{\'e} J, Wang J, M{\"u}ller J, Spiecker E, Mills MJ,
  Povstugar I, Choi P, Raabe D, Bitzek E (2015) Atom probe informed simulations
  of dislocation--precipitate interactions reveal the importance of local
  interface curvature, Acta Mater 92:33 -- 45

\bibitem{Krug-2014}
Krug ME, Mao Z, Seidman DN, Dunand DC (2014) A dislocation dynamics model of
  the yield stress employing experimentally-derived precipitate fields in
  {Al-Li-Sc} alloys, Acta Mater 79:382--395

\bibitem{pareige2011kinetic}
Pareige C, Roussel M, Novy S, Kuksenko V, Olsson P, Domain C, Pareige P (2011)
  Kinetic study of phase transformation in a highly concentrated fe--cr alloy:
  {M}onte {C}arlo simulation versus experiments, Acta Mater 59:2404--2411

\bibitem{Moody-2014}
Moody MP, Ceguerra AV, Breen AJ, Cui XY, Gault B, Stephenson LT, Marceau RKW,
  Powles RC, Ringer SP (2014) Atomically resolved tomography to directly inform
  simulations for structure--property relationships, Nat Commun 5

\bibitem{Uchic2004}
Uchic MD, Dimiduk DM, Florando JN, Nix WD (2004) Sample dimensions influence
  strength and crystal plasticity, Science 305:986--989

\bibitem{Broderick2008}
Broderick S, Suh C, Nowers J, Vogel B, Mallapragada S, Narasimhan B, Rajan K
  (2008) Informatics for combinatorial materials science, {JOM} 60:56--59

\bibitem{Rajan2002}
Rajan K, Rajagopalan A, Suh C (2002) Data mining and multivariate analysis in
  materials science, in: Gaune-Escard M (ed.) {Molten Salts: From Fundamentals
  to Applications}, Springer Netherlands, no.~52 in {NATO} {Science} {Series},
  pp. 241--248

\bibitem{TMSMultiscale}
Voorhees P, Spanos G, et~al. (2015) Modeling across scales: {A} roadmapping
  study for connecting materials models and simulations across length and time
  scales, Tech. rep., The Minerals, Metals and Materials Society (TMS)

\bibitem{JuulJensen2008}
Juul~Jensen D, Godiksen RBN (2008) Neutron and synchrotron {X-ray} studies of
  recrystallization kinetics, Metall Mater Trans A 39A:3065--3069

\bibitem{Babinsky2014}
Babinsky K, De~Kloe R, Clemens H, Primig S (2014) A novel approach for
  site-specific atom probe specimen preparation by focused ion beam and
  transmission electron backscatter diffraction, Ultramicroscopy {144}:9--18

\bibitem{juan2012prediction}
Juan PA, Berbenni S, Capolungo L (2012) Prediction of internal stresses during
  growth of first-and second-generation twins in {Mg} and {Mg} alloys, Acta
  Mater 60:476--486

\bibitem{abdolvand2015study}
Abdolvand H, Majkut M, Oddershede J, Wright JP, Daymond MR (2015) Study of
  3-{D} stress development in parent and twin pairs of a hexagonal close-packed
  polycrystal: {Part II}--crystal plasticity finite element modeling, Acta
  Mater 93:235--245

\bibitem{woo1992production}
Woo C, Singh B (1992) Production bias due to clustering of point defects in
  irradiation-induced cascades, Phil Mag A 65:889--912

\bibitem{Marsden-2014}
Marsden W, Cebon D, Cope E (2014) Managing multi-scale material data for access
  within {ICME} environments, in: Arnold SM, Wong TT (eds.) Models, Databases
  and Simulation Tools Needed for Realization of Integrated Computational
  Materials Engineering, ASM International, pp. 82--90

\end{thebibliography}
\end{document}